\documentstyle[prb,preprint,tighten,aps,eqsecnum]{revtex}

\begin{document}
\draft
\title{The periodic Anderson model from the atomic limit and FeSi}
\author{M. E. Foglio}
\address{Instituto de F\'{\i}sica ``Gleb Wataghin''\\
Universidade Estadual de Campinas,UNICAMP\\
13083-970 Campinas, S\~{a}o Paulo, Brasil}
\author{M. S. Figueira}
\address{Instituto de F\'{\i}sica, C.P. 100.093\\
Universidade Federal Fluminense,UFF\\
24001-970 Niter\'{o}i, Rio de Janeiro, Brasil}
\date{\today}
\maketitle

\begin{abstract}
The exact Green's functions of the periodic Anderson model for $U\rightarrow
\infty $ are formally expressed within the cumulant expansion in terms of an
effective cumulant. Here we resort to a calculation in which this quantity
is approximated by the value it takes for the exactly soluble atomic limit
of the same model. In the Kondo region a spectral density is obtained that
shows near the Fermi surface a structure with the properties of the Kondo
peak. Approximate expressions are obtained for the static conductivity $%
\sigma (T)$ and magnetic susceptibility $\chi (T)$ of the PAM, and they are
employed to fit the experimental values of FeSi, a compound that behaves like 
a Kondo insulator with both quantities vanishing rapidly for $T\rightarrow 0$.
Assuming that the system is in the intermediate valence region, it was
possible to find good agreement between theory and experiment for these two
properties by employing the same set of parameters. It is shown that in the 
present model the hybridization is responsible for the relaxation mechanism 
of the conduction electrons.
\end{abstract}





\section{Introduction}

\label{S01}

In the present work we discuss approximate Green's Functions (GF) of the
Periodic Anderson Model (PAM) that use the atomic limit as a starting point.
We employ these GF to calculate the static magnetic susceptibility and the
resistivity of {\rm FeSi}.

{\rm FeSi} has rather unusual magnetic properties,\cite{Jacar} and in
particular a static susceptibility $\chi (T)$ that has a maximum $\chi
(T_{m})$ at about $T_{m}=536$ K and vanishes for $T\rightarrow 0$ when the
low temperature Curie tail is subtracted.\cite{Schles} Several models were
studied by Jacarino et.al.,\cite{Jacar} and they found that a simple one,
with two very narrow rectangular bands separated by a gap, could be used to
fit $\chi (T)$. The properties of FeSi are very similar to those of the
Kondo insulators,\cite{Schles,Aeppli,Schles2} and a very simple model that
describes most of their properties would consist of two hybridized bands
with two electrons,\cite{Aeppli} i.e.: an intrinsic semiconductor with an
hybridization gap.\ Both $\chi (T)$ and the resistivity $\rho (T)$ of
FeSi have activation laws with characteristic energies of order 0.1 eV, and
although band structure calculations\cite{Mattheis} give comparable
semiconducting gaps they can not reproduce the large values of $\chi (T)$.
Also the measurements of infrared and optical reflectivity can not be
described with the predictions of the band calculations,\cite{Schles2} and
the PAM seems a more adequate model, both because it becomes the previous
model when the Coulomb repulsion between the localized electrons is
neglected, i.e.: $U=0$, and because the correlations present in the PAM
could explain some of the observed features of FeSi.

A two-band Hubbard model\cite{Fu} has been already employed to study FeSi,
and it is equal to the PAM when the dispersion in one band and the Coulomb
repulsion in the other are zeroed. Another variant of the PAM was to add
dispersion to the band of localized electrons, and this model has been
studied both with $U\rightarrow \infty $\cite{Mucio} and with finite U.\cite
{TovarTOJC} It has been suggested\cite{Aeppli} that it would be interesting
to describe the Kondo insulators employing the PAM with $U\rightarrow \infty 
$, and this approach is presented here to calculate both $\chi (T)$ and $%
\rho (T)$ employing approximate\cite{Foglio} GF. This model has been
also used to study the Kondo insulator ${\rm Ce}_{3}{\rm Bi}_{4}{\rm Pt}_{3}$%
, employing the slave boson technique in the mean field approximation.\cite
{SanchezBC,Riseborough} Varma \cite{Varma} argues that the strict Kondo
lattice (i.e. with integer f occupation and therefore without charge
fluctuations) is inappropriate for these systems, because it would be rather
unlikely to find the chemical potential just in the hybridization gap, and
that one would more likely find this situation in a mixed valence system.
Our calculation shows that with the simplified model presented here, it is
possible to give a fair description of the $\chi (T)$ and $\rho (T)$ of
FeSi in a typical intermediate valence situation.\cite{Note1}

In the study of solid state systems it is sometimes interesting to focus on
the local states of the ions placed at the different sites of the crystal 
\cite{Foglio}. The space of the local or ionic states associated to a given
site contains many states that are usually of little interest, generally
because their occupation can be neglected at fairly low temperatures or
frequencies when they are too far apart in energy from the ground state. It
is then useful to eliminate these states from the model of the system, and
the Hubbard operators\cite{Hubbard4,HaleyErdos} are very convenient for that
purpose. In the Anderson lattice we have a broad band of conduction
electrons, identified by a subindex $c,$ and the local $d$ states, which
will be identified with the subindex $f$ for convenience. At each site $j$
of the lattice there are four local states: the vacuum state \ $\left| \
j,0\right\rangle $, the two states $\left| \ j,\sigma \right\rangle $ of one
electron with spin component $\sigma $ and the state $\left| \
j,2\right\rangle $ with two local electrons. \ When $U\rightarrow \infty $ \
the state $\left| \ j,2\right\rangle $ is empty, and one can use the Hubbard
operators to project it out from the space of local states at site $j$. One
difficulty is that the usual expansions employed with the Fermi or Bose
operators are not valid for these operators, and Hubbard introduced for his
model of correlated electrons~\cite{Hubbard123} a diagrammatic expansion
with cumulants~\cite{Hubbard5} which uses the electron hopping as a
perturbation, and becomes the usual expansion when $U=0$. The GFs employed
in the present work are based on an extension \cite{FFM} of Hubbard's
cumulant expansion that is valid for the Anderson lattice and uses the
hybridization as perturbation; employing this expansion it is possible to
express the exact GF in terms of an unknown effective cumulant\cite
{Foglio,Metzner} $M_{2,\sigma }^{eff}(\omega )$. In this work we shall
approximate the effective cumulant by $M_{2,\sigma }^{at}(z)$, which is
obtained from the exact solution of the Anderson lattice in the atomic
limit, namely when the band of uncorrelated electrons has zero width\cite
{Foglio}. The spectral density obtained in this approximation, for typical
values of the system parameters, shows a structure close to the chemical
potential $\mu $ that corresponds to the Kondo resonance and affects the
physical properties at low temperatures. This structure was absent from the
GF derived from the cumulant expansion when only cumulants up to fourth
order were employed\cite{FFM,Figueira}, and this was the motivation to
approximate the effective cumulant by a method that would include all the
higher order cumulants that were absent in our previous calculations.
Because of its atomic character, the approximate effective cumulant in our
method is independent of the wave vector. Our GFs are therefore closely
related to those employed by the dynamical mean field theory\cite{Kraut} of
the infinite dimensional problem,\cite{MetznerV,MullerH} but rather than
using a self consistency condition we use physical considerations to chose
the parameters that define the effective cumulant. The dynamical mean field
theory has been also employed to study the magnetism of the PAM\cite
{Rozenberg} as well as the transfer of spectral weight in the spectroscopy
of correlated electron systems described by this model.\cite{RozenbergKK}

Both the dynamical susceptibility and the conductivity require two-particle
GFs in the theory of linear response, but we shall obtain approximate
expressions of the static quantities employing the one-particle GFs
introduced in this work. With these GF we obtain the total number of spin up
and of spin down electrons in the presence of a weak magnetic field, and the
static susceptibility $\chi (T)$ is then proportional to their difference
divided into the magnetic field; we can then compare the ratio $\chi
(T)/\chi (T_{m})$ with the corresponding experimental value. To simplify the
calculation of the effective cumulant we assume equal gyromagnetic factors $%
g_{f}$ \ and $g_{c}$, although the extension to different $g_{f}$ \ and $%
g_{c}$ would not present essential difficulties. 

It has been shown \cite{SchweitzerC} that the limit $d=\infty $ can be used
to give an approximate\ description of three-dimensional systems, and we
shall then use an expression of $\sigma \left( T\right) $ that is valid
in infinite dimension in our calculation for {\rm FeSi}. This expression is
derived from the well known Kubo formula, the vertex corrections cancel out
when $d=\infty ,$\cite{Khurana} and only the one-particle $G_{c,\sigma }(%
{\bf k},z)$ are then necessary to calculate $\sigma \left(T\right) $. 
In this expression there are explicit sums over ${\bf k}$, but
when nearest-neighbor hopping in a simple cubic lattice is considered, it is
possible to derive expressions\cite{MutouH,PruschkeCJ} that depend on ${\bf k%
}$ only through the unperturbed conduction electron energies $\varepsilon
\left( {\bf k}\right) $. As a further simplification that would not change
the results in an essential way, we shall use a rectangular band  with $%
-W\leq \varepsilon \left( {\bf k}\right) $ $\leq W$.

Using the expressions of $\sigma \left(T \right) $ and $\chi (T)$
discussed above we fitted the experimental magnetic susceptibility \cite
{Jacar} $\chi (T)/\chi (T_{m})$ and the static resistivity\cite{Schles} $%
\rho \left( T\right) =1/\sigma \left( T\right) $ of {\rm FeSi}\ with the
same set of parameters in a typical situation of intermediate valence,
obtaining a fairly reasonable agreement with the experimental values. To
adjust the $\chi (T)$ at high $T$, it was necessary to assume that the
thermal expansion affects the value of the system's parameters.

In Section \ref{S02} we discuss the Periodic Anderson Model (PAM) and the
approximate one-particle GF employed in this work. In Section \ref{S03} we
analyze the static magnetic susceptibility $\chi (T)$ and the static
resistivity $\rho \left(T\right) $ of FeSi. Conclusions are presented in
Section \ref{S04}.



\section{Green's functions for the Periodic Anderson model}

\label{S02}

As discussed in Section \ref{S01} we employ Hubbard's operators.\cite
{Hubbard4} In the general case there is a fixed number $n$ of orthogonal
states $\{\left| j,a\right\rangle \}$ (identified by indices $a$) that span
a space ${\cal E}_{j,n}$ at each site $j$, where $j=1,2,\ldots ,N_{s}$ and $%
N_{s}$ is the number of sites. To each site $j$ we associate the $n^{2}$
Hubbard operators 
\begin{equation}
X_{j,ab}=\left| j,a\right\rangle \left\langle j,b\right| \qquad ,
\label{E2.1}
\end{equation}
which transform the state $\left| j,b\right\rangle $ into the state $\left|
j,a\right\rangle ,$ i.e. $X_{j,ab}\left| j,b\right\rangle =\left|
j,a\right\rangle $. The product rules for two operators at the same site are
given by 
\begin{equation}
X_{j,ab}\ X_{j,cd}=\delta _{b,c}\ X_{j,ad}\qquad ,  \label{E2.2}
\end{equation}
and we chose properties equivalent to those of the usual Fermi or Bose
operators when they are at different sites. We then say that $X_{j,ab}$ is
of the ``Fermi type'' (``Bose type'') when the number of electrons in the
two states $\mid j,a\rangle $ and $\mid j,b\rangle $ differ by an odd (even)
number. For $j\neq j^{\prime }$ we then use $\left\{ X_{j,ab},X_{j^{\prime
},cd}\right\} =0$ when the two operators are of the ``Fermi type'' and $%
\left[ X_{j,ab},X_{j^{\prime },cd}\right] =0$ when at least one is of the
``Bose type'' (as usual\cite{FetterW} $[a,b]=ab-ba$ and $\{a,b\}=ab+ba$).

\subsection{The Anderson lattice for $U\rightarrow \infty $ \label{S02_1}}

The Anderson lattice with finite $U$ is described by the Hamiltonian 
\begin{eqnarray}
H &=&\sum_{{\bf k}\sigma }\ E_{{\bf k},\sigma }\ C_{{\bf k}\sigma }^{\dagger
}C_{{\bf k}\sigma }+\sum_{j\sigma }E_{j,\sigma }\ f_{j\sigma }^{\dagger
}f_{j\sigma }+\sum_{j}U\ f_{j\sigma }^{\dagger }f_{j\sigma }f_{j\overline{%
\sigma }}^{\dagger }f_{j\overline{\sigma }}+  \nonumber \\
&&\sum_{jk\sigma }\left( V_{j,{\bf k},\sigma }\ f_{j\sigma }^{\dagger }C_{%
{\bf k}\sigma }\ +H.C.\right) \ ,\   \label{E2.3}
\end{eqnarray}
where $C_{k\sigma }^{\dagger }$ ($C_{k\sigma }$) is the usual creation
(annihilation) operator of conduction band electrons with wavevector ${\bf k}
$ and spin component $\sigma \hbar /2$, where $\sigma =\pm 1$. The $%
f_{j\sigma }^{\dagger }$ $\ $and $f_{j\sigma }$ correspond to the local f or
d electrons at site $j$ , and 
\begin{equation}
V_{j,{\bf k},\sigma }=V(k)\exp \left( i{\bf k.{R}_{j}}\right) \qquad .
\label{E2.3a}
\end{equation}
When $U=0$ Eq.(\ref{E2.3}) describes two hybridized bands of uncorrelated
electrons.

The state space of the f-electrons at each site $j$ is spanned by the four
states$\mid j,0\rangle ,$ $\mid j,\sigma \rangle $ and $\mid j,2\rangle $,
with $\sigma =\pm 1$. The state $\mid j,2\rangle $ is empty when $%
U\rightarrow \infty $, and we shall consider a reduced space of states by
projecting $\mid j,2\rangle $ out, so that ${\cal E}_{j,n}$ is a three
dimensional space. To make the connection with the Hubbard operators one
could substitute the identity\cite{identity} 
\begin{equation}
f_{j\sigma }=X_{j,0\sigma }+\sigma X_{j,\overline{\sigma }d}  \label{E2.4}
\end{equation}
into Eq. (\ref{E2.3}), where the factor $\sigma $ is necessary to obtain the
correct phase of the states. Eliminating the $X_{j,\overline{\sigma }d}$ and 
$X_{j,22}$ one obtains the projection of $H$ into the reduced space, namely 
\begin{eqnarray}
H_{r} &=&\sum_{{\bf k}\sigma }\ E_{{\bf k},\sigma }\ C_{{\bf k},\sigma
}^{\dagger }C_{{\bf k},\sigma }+\sum_{j,\sigma }\ E_{j,\sigma }\ X_{j,\sigma
\sigma }+  \nonumber \\
&&\sum_{j{\bf k}\sigma }\left( V_{j,{\bf k},\sigma }\ X_{j,0\sigma
}^{\dagger }\ C_{{\bf k},\sigma }+V_{j,{\bf k},\sigma }^{\ast }\ C_{{\bf k}%
,\sigma }^{\dagger }\ X_{j,0\sigma }\right) \qquad .  \label{E2.5}
\end{eqnarray}

\subsubsection{The cumulant expansion\label{S02_2}}

The cumulant expansion has been employed by several authors to study Ising's
and Heisenberg's models,\cite{Wortis} while Hubbard\cite{Hubbard5} extended
the method to a quantum problem with fermions. In this technique the
cumulant averages\cite{Kubo} are used to rearrange the usual perturbative
expansion,\cite{Hubbard5} and it is possible to derive a diagrammatic
expansion involving unrestricted lattice sums of connected diagrams, that
satisfies a linked cluster theorem. This technique was extended to the
Anderson lattice,\cite{FFM} and a brief description is given here. The
method employs the Grand Canonical Ensemble of electrons, and it is then
convenient to introduce 
\begin{equation}
{\cal H}=H-\mu \left\{ \sum_{\vec{k},\sigma }C_{\vec{k},\sigma }^{\dagger
}C_{\vec{k},\sigma }+\sum_{ja}\nu _{a}X_{j,aa}\right\} \qquad ,  \label{E2.6}
\end{equation}

\noindent where $\mu $ is the chemical potential and $\nu _{a}$ is the
number of electrons in the state $\left| j,a\right\rangle $. The last term
in Eq. (\ref{E2.5}) will be considered as the perturbation, and the exact
and unperturbed averages of any operator $A$ are respectively denoted by $%
<A>_{{\cal H}}$ and $<A>$. It is also convenient to introduce 
\begin{eqnarray}
\varepsilon _{j,a} &=&E_{j,a}-\mu \ \nu _{a}  \label{E2.6a} \\
\varepsilon _{{\bf k\sigma }} &=&E_{{\bf k\sigma }}-\mu \ \qquad ,
\label{E2.6b}
\end{eqnarray}
because these are the forms that consistently appear in the calculations.

The Matsubara expansion\cite{FetterW} is employed, so that $\tau $ is an
imaginary time in the GFs $\left\langle \left( \widehat{X}_{j,\alpha }(\tau
)\ \widehat{X}_{j^{\prime },\alpha ^{\prime }}(\tau ^{\prime })\right)
_{+}\right\rangle _{{\cal H}}$, where $\alpha \equiv (a,b)$ identifies the
transition $b\rightarrow a$, and 
\begin{equation}
\widehat{X}_{j,\alpha }(\tau )=\exp \left( \tau {\cal H}\right) X_{j,\alpha
}\exp \left( -\tau {\cal H}\right) \qquad .  \label{E2.7}
\end{equation}
The subindex $+$ in the definition of the GF indicates that the operators
inside the parenthesis are taken in the order of increasing $\tau $ to the
left, with a change of sign when two Fermi-type operators have to be
exchanged to obtain this ordering. The inverse of Plank's $\hbar $ and of
Boltzmann's $k_{B}$ are usually included into the real and imaginary 
temperatures $T$ and $\tau $, as well as in several other
parameters, so that all of them are given in terms of a common energy unit.

Some of the infinite diagrams which contribute to the GF $\left\langle
\left( \widehat{X}_{j,\alpha }(\tau )\ \widehat{X}_{j^{\prime },\alpha
^{\prime }}(\tau ^{\prime })\right) _{+}\right\rangle _{{\cal H}}$ are shown
in figure~\ref{F4_1}, and the full circles (f-vertices) correspond to the
cumulants of the f-electrons. Each line reaching a vertex is associated to
one of the $X$ operators of the cumulant, and the free lines (i.e. those
that do not join an empty circle) correspond to the external $X$ operators
appearing in the exact GF. An explicit definition of the cumulants can be
found in the references \onlinecite{Hubbard5,FFM,FFM2}, and they can be
calculated by employing a generalized Wick's theorem.\cite{FFM2,Hewson,YangW}

The first diagram in figure~\ref{F4_1}a corresponds to the simplest free
propagator $\left\langle \left( X_{j,\alpha }(\tau )\ X_{j^{\prime },\alpha
^{\prime }}\right) _{+}\right\rangle $, and the second diagram in that
figure has an empty circle (c-vertex) that corresponds to the conduction
electron cumulant, equal to the free propagator$\left\langle \left(
C_{k\sigma }(\tau )\ C_{k\sigma }^{\dagger }\right) _{+}\right\rangle \equiv 
$ $G_{c{,}\sigma }^{o}({\bf k}{,}\tau )$. The interaction is represented by
the lines (edges) joining two vertices and, because of the structure of the
hybridization, they always join a c-vertex to an f-vertex; the number of
edges in a diagram gives its order in the perturbation expansion.

Cumulants containing statistically independent operators are zero, and those
appearing in the present formalism (with the hybridization as perturbation)
vanish unless they contain only $X$ operators at the same site or only $C$
or $C^{\dagger }$ operators with the same $k$ and $\sigma $. The only
non-zero c-cumulants are of second order, because the uncorrelated
c-operators satisfy Wick's theorem. On the other hand, the f-vertices can
have many legs, all corresponding to $X$ operators at the same site, like
the fourth and sixth order cumulants appearing in the rather more
complicated diagram shown in figure~\ref{F4_1}c.

All the infinite diagrams that contribute to the GF with cumulants of at
most second order are shown in figure~\ref{F4_1}a, and this family is the
``chain approximation'' (CHA), which gives the exact solution of Eq. (\ref
{E2.3}) when there is no Coulomb correlation ($U=0$). When the spin is
eliminated from the problem, the Hamiltonian of equation~(\ref{E2.5})
corresponds to a system of two hybridized bands without any Coulomb
repulsion (there can be only one or zero f-electrons at each site), and the
CHA is again an exact solution.\cite{FFM} The diagrams of the c-electron
propagator in the same approximation are shown in figure \ref{F4_1}b.

In the Feynmann perturbation expansion, Wick's theorem is valid and only
second order propagators appear, while the interactions are provided by the
Coulomb interaction. In the present treatment, the $U$ disappears in the
limit $U\rightarrow \infty $ (or is included in the unperturbed Hamiltonian
when $U$ is finite), and the correlations appear through the non-zero
cumulants of $X$ operators with order greater than two, which include
propagators of two or more particles. In the Feynmann expansion of the one
particle GF, the two particle GF appear in the self-energy, which contains
all the correlations.

\subsubsection{The spectral density of the GF and the occupation number\label%
{S02-3}}

In the Anderson lattice with $U\rightarrow \infty $ one can introduce
one-particle GFs of local electrons 
\begin{equation}
\left\langle \left( X_{j,0\sigma }(\tau )\ X_{j^{\prime },\sigma 0}(\tau
^{\prime })\right) _{+}\right\rangle _{{\cal H}}\qquad ,  \label{E4.8}
\end{equation}
as well as GFs\ for the c-electrons $\left\langle \left( C_{k\sigma }(\tau
)\ C_{k^{\prime }\sigma }^{\dagger }(\tau ^{\prime })\right)
_{+}\right\rangle _{{\cal H}}$\ and ``crossed'' GFs of the type $%
\left\langle \left( X_{j,0\sigma }(\tau )\ C_{k^{\prime }\sigma }^{\dagger
}(\tau ^{\prime })\right) _{+}\right\rangle _{{\cal H}}$, all of them
defined in the intervals $0\leq \tau ,\tau ^{\prime }\leq \beta \equiv 1/T$.
It is possible to associate a Fourier series to these GFs because of their
boundary condition in this variable,\cite{FFM} and the coefficients
correspond to the Matsubara frequencies $\omega _{\nu }=\pi \nu /\beta $
(where $\nu $ are all the positive and negative odd integer numbers). One
can also transform the GF to reciprocal space\cite{FFM} so that 
\begin{eqnarray}
\left\langle \left( X_{j,0\sigma }(\tau )\ X_{j^{\prime },\sigma 0}(\tau
^{\prime })\right) _{+}\right\rangle _{{\cal H}} &=&\frac{1}{\beta \ N_{s}}%
\sum_{{\bf k},{\bf k}^{\prime }}\sum_{\omega _{\nu },\omega _{\nu ^{\prime
}}}\ \exp \left[ i\left( {\bf k}.{\bf R}_{j}-{\bf k}^{\prime }.{\bf R}%
_{j}^{\prime }\right) \right.  \nonumber \\
&&\left. -i\left( \omega _{\nu }\ \tau +\omega _{\nu ^{\prime }}\ \tau
^{\prime }\right) \right] \ \left\langle \left( X_{{\bf k},0\sigma }(\omega
_{\nu })\ X_{{\bf k}^{\prime },\sigma 0}(\omega _{\nu ^{\prime }})\right)
_{+}\right\rangle _{{\cal H}}\qquad ,  \label{E4.8a}
\end{eqnarray}
and because of the invariance against time and lattice translations\cite
{FFM,FF2} 
\begin{equation}
\left\langle \left( X_{{\bf k},0\sigma }(\omega _{\nu })\ X_{{\bf k}^{\prime
},\sigma 0}(\omega _{\nu ^{\prime }})\right) _{+}\right\rangle _{{\cal H}%
}=G_{ff,\sigma }({\bf k},\omega _{\nu })\ \delta _{{\bf k}^{\prime },{\bf k}%
}\ \delta _{\nu +{\nu }^{\prime },0}\qquad ,  \label{E4.8b}
\end{equation}
where $\delta _{{\bf k}^{\prime },{\bf k}}$ and $\delta _{\nu +\nu ^{\prime
},0}$ are Kronecker's deltas, because the ${\bf k}$ and $\omega _{\nu }$ are
discrete variables. Transforming the eigenstates of the c-electrons to the
Wannier representation, one obtains the equivalent relations for $%
G_{cc,\sigma }({\bf k},\omega _{\nu })$ and $G_{fc,\sigma }({\bf k},\omega
_{\nu })$. Considering that the coefficients of the $\tau $ Fourier series
for each ${\bf k}$ are the values of a function of the complex variable $%
z=\omega +iy$ at the points $z_{\nu }=i\ \omega _{\nu }$, it is possible to
make the analytic continuation to the upper and lower half-planes of $z$ in
the usual way,\cite{Negele} obtaining, e.g. from the $G_{ff,\sigma }({\bf k}%
,\omega _{\nu })$, a function $G_{ff,\sigma }({\bf k},z)$ which is minus the
Fourier transform of the double time GF.\cite{Zubarev}

If we assume the system to be uniform, the occupation of the local state $%
\left\langle X_{j,\sigma \sigma }\right\rangle $ does not depend on $j$, and
it is given by 
\begin{equation}
n_{f,\sigma }=\int_{-\infty }^{\infty }\rho _{f,0\sigma }(\omega )\
f_{T}(\omega )\ d\omega \qquad ,  \label{E4.9}
\end{equation}

\noindent where 
\begin{equation}
f_{T}(z)=\left[ 1+\exp \left( \beta \ z\right) \right] ^{-1}  \label{E4.10}
\end{equation}
is the Fermi function and 
\begin{equation}
\rho _{f,0\sigma }(\omega )=\frac{1}{\pi }\ \lim_{\eta \rightarrow 0}\ {\it %
Im}\left\{ \frac{1}{N_{s}}\sum_{{\bf k}}G_{ff,\sigma }({\bf k},\omega
+i\left| \eta \right| )\right\}  \label{E4.11}
\end{equation}
is the spectral density associated to the transition $\sigma \rightarrow 0$,
abbreviated with $(0,\sigma )$. Using the same $\rho _{f,0\sigma }(\omega )$
it is also possible to obtain the occupation $\left\langle
X_{j,00}\right\rangle =n_{f,0}$ of the empty state $\mid 0\rangle $, namely 
\begin{equation}
n_{f,0}=\int_{-\infty }^{\infty }\rho _{f,0\sigma }(\omega )\
(1-f_{T}(\omega ))\ d\omega \qquad .  \label{E4.12}
\end{equation}
All the corresponding quantities for the conduction electrons are obtained
by replacing $G_{cc,\sigma }({\bf k},z)$ for $G_{ff,\sigma }({\bf k},z)$.

The \ f-electron GF is given in the CHA (cf. the diagrams in figure \ref
{F4_1}a) by 
\begin{equation}
G_{ff,\sigma }({\bf k},z)=-\frac{D_{\sigma }^{0}(z-\varepsilon _{{\bf k}%
\sigma })}{(z-\varepsilon _{1,\sigma }({\bf k}))(z-\varepsilon _{2,\sigma }(%
{\bf {k}}))}\qquad ,  \label{4.13}
\end{equation}
where the energies $\varepsilon _{1,\sigma }({\bf k})$ and $\varepsilon
_{2,\sigma }({\bf k})$ are the two elementary excitations with wave vector $%
{\bf k}$ and spin component $\sigma $, resulting from the hybridization of a
band $\varepsilon _{{\bf k}\sigma }$ and a dispersionless band of energy $%
\varepsilon _{f}=E_{{j}\sigma }-\mu $, with a reduced hybridization constant 
$\sqrt{D_{\sigma }}V(k)$. These energies are given by the two roots of $%
(z-\varepsilon _{f})(z-\varepsilon _{{\bf k}\sigma })-D_{\sigma }\left|
V(k)\right| ^{2}=0$, where 
\begin{equation}
D_{\sigma }=\int_{-\infty }^{\infty }\rho _{f,0\sigma }(\omega )\ d\omega
=\left\langle X_{00}+X_{\sigma \sigma }\right\rangle \qquad .  \label{E4.14}
\end{equation}

The spectral density for the unperturbed f-electrons is a $\delta $-function
at $\varepsilon _{f}$, and $\rho _{f,0\sigma }(\omega )$ becomes two bands
with a gap centered at $\varepsilon _{f}$ and roughly proportional to the
reduced hybridization constant $\sqrt{D_{\sigma }}V$ in the CHA. When the
system is in the ``Kondo region'' in which the local state has nearly the
maximum occupation compatible with the paramagnetic state, viz. $n_{f,\sigma
}=0.5$ in the average, a narrow temperature-dependent peak,\thinspace the
``Kondo peak'', should appear very close to the chemical potential, i.e.
near to $\omega =0$ in the variables we use. This peak is responsible for
many properties of the ``heavy fermions'',\cite{Reviews} and it is closely
related to the increase in resistivity when T decreases, observed in many
metals with magnetic impurities at low T. In combination with the effect of
phonons, that make the resistivity increase with T, it is responsible for
the minimum in resistivity, named the Kondo effect, that is observed in
those systems. To explain this behavior it was essential to consider a third
order perturbation that includes spin-flip processes.\cite{Kondo} These
processes are absent from the CHA, because diagrams with only second order
cumulants do not include them, so that the absence of the Kondo peak in that
approximation is not surprising. The GFs proposed in this work avoid the
laborious explicit calculation of higher order cumulants by including all of
them in an approximate way.

\subsubsection{The exact Green's functions\label{S02_4}}

In the calculation with the usual Fermi or Bose operators, the one-particle
propagator of the f-electron is given by a sum of diagrams of the same type 
\cite{Note01} shown in figure \ref{F4_1}a but with each vertex corresponding
to the sum of all ``proper'' (or irreducible) diagrams.\cite
{FetterW,LuttingerW} The same result is found in the cumulant expansion of
the Hubbard model for $d\rightarrow \infty $,\cite{Metzner,CracoG} when the
electron hopping is employed as perturbation. The vertices then represent an
``effective cumulant'' $M_{2,\sigma }^{eff}(z)$, that is independent of $%
{\bf k}$ because only diagrams of a special type contribute to this quantity
for $d\rightarrow \infty $.

In the cumulant expansion of the Anderson lattice\cite{FFM} we employ the
hybridization rather than the hopping as a perturbation, and the exact
solution of the conduction electrons problem in the absence of hybridization
is part of the zeroth order Hamiltonian. For this reason it became necessary
to extend Metzner's derivation\cite{Metzner} to the Anderson lattice, and we
have shown\cite{FFdi} that the same type of results obtained by Metzner are
also valid for this model. These results had been used\cite{Foglio} to
obtain the exact GF employed in the present work, but since then we realized
that the expression of the exact GF is valid for all dimensions and it is
not necessary to assume infinite dimension in that part of the derivation.
As with the Feynmann diagrams, one can rearrange all those that contribute
to the exact $G_{ff,\sigma }({\bf k},z)$ by defining an effective cumulant $%
M_{2,\sigma }^{eff}({\bf k},z)$, that is given by all the diagrams of $%
G_{ff,\sigma }({\bf k},z)$ that can not be separated by cutting a single
edge (usually called ``proper'' or ``irreducible'' diagrams). The exact GF $%
G_{ff,\sigma }({\bf k},z)$ is then given by the family of diagrams in figure 
\ref{F4_1}a, but with the effective cumulant $M_{2,\sigma }^{eff}({\bf k,}z)$
in place of the bare cumulant $M_{2,\sigma }^{0}(z)=-D_{\sigma
}^{0}/(z-\varepsilon _{f})$ at all the filled vertices.The exact GF for the
f electron is then written as 
\begin{equation}
G_{ff,\sigma }({\bf k},z)=M_{2,\sigma }^{eff}({\bf k}{,}z)\ \frac{1}{1-\mid
V({\bf k})\mid ^{2}G_{c{,}\sigma }^{o}({\bf k}{,}z)\ M_{2,\sigma }^{eff}(%
{\bf k}{,}z)}\qquad ,  \label{E2.8}
\end{equation}
where $G_{c{,}\sigma }^{o}({\bf k}{,}z)=-1/(z-\varepsilon ({\bf k}))$ is the
frequency Fourier transform of $G_{c{,}\sigma }^{o}({\bf k}{,\tau })$, and
in a similar way one obtains the exact GF for the c-electron, namely 
\begin{equation}
G_{cc,\sigma }({\bf k},z)=\frac{-\ 1}{z-\varepsilon ({\bf k})+\mid V({\bf k}%
)\mid ^{2}M_{2,\sigma }^{eff}({\bf k}{,}z)}\qquad .  \label{E2.8b}
\end{equation}

It is clear that for arbitrary dimension we have not gained much with Eq.(%
\ref{E2.8}), because the calculation of $\ M_{2,\sigma }^{eff}({\bf k}{,}z)$
is as difficult as that of $G_{ff,\sigma }({\bf k},z)$. We circumvent this
obstacle by resorting to replace $M_{2,\sigma }^{eff}({\bf k}{,}z)$ by the
corresponding quantity $M_{2,\sigma }^{at}(z)$ of an exactly soluble model,
which is the atomic limit of the Anderson lattice. The hopping is neglected
in this system, described by the Hamiltonian of Eqs. (\ref{E2.3}) or (\ref
{E2.5}) with $E_{{\bf k},\sigma }=E_{0}$, and it already contains the basic
physics of the formation of the singlet ground state and of the appearance
of the Kondo peak in the PAM, as it is discussed in review articles.\cite
{Atomic} Because of its atomic character, the approximate effective cumulant 
$M_{2,\sigma }^{at}(z)$ thus obtained is independent of ${\bf k}$,\ and can
be calculated exactly as shown in the next section. Being a special case of
the PAM, it implicitly contains all the higher order cumulants.

The effective cumulant is also independent of ${\bf k}$ when $d\rightarrow
\infty $, and the very successful ``dynamical mean field theory''\cite{Kraut}
also employs GFs corresponding to those in Eqs.~(\ref{E2.8},\ref{E2.8b}).

\subsection{The exact solution in the atomic limit}

The exact solution of the local problem has been already used in different
ways to study the Anderson lattice. The limit $U\rightarrow \infty $ was
studied in the intermediate valence case\cite{FoglioF} by considering only
the lowest four eigenstates of the local Hamiltonian and the magnetic
instabilities and susceptibility were discussed employing the resulting
self-consistent Hamiltonian.\cite{FoglioBF} Considering only the atomic
limit, viz. taking $E_{{\bf k},\sigma }=E_{0}^{a}$, Alascio et.al.\cite
{AlascioAA,AlascioAA1} studied the model for the whole range of parameters,
showing that ``most of the essential characteristics'' of these systems
``are present in this crudely simplified Hamiltonian''. Sim\~{o}es et.al. 
\cite{SimoesIA} employed the atomic limit together with a diagrammatic
method,\cite{Anda} that is essentially equivalent to our CHA, considering
both the hopping and the hybridization as perturbations. An important
improvement of the technique was to apply the same diagrammatic expansion to
the exact solution of the atomic limit,\cite{SimoesIRA,Simoes} employing
only the hopping as perturbation, and this technique has also been applied
to study the problem with finite $U$.\cite{BrunetRI,BrunetGI}

As discussed in the previous section we introduce the exact expression for
the GF that is given in terms of an effective cumulant $M_{2,\sigma }^{eff}(%
{\bf k},z)$, however replacing this quantity by the approximate $M_{2,\sigma
}^{at}(z)$. As this treatment was derived from the diagrammatic expansion 
\cite{FFM} which uses the hybridization as perturbation and employs the
exact solution of the uncorrelated conduction band, it seems a better
starting point than considering the hopping as a perturbation, because the
hybridization is usually rather smaller than the bandwidth. The atomic limit
has also been employed, within the framework of the dynamical mean field
theory, to study the transport properties of the symmetric PAM.\cite
{Consiglio}

\subsubsection{The exact GF in the atomic limit}

Taking $E_{{\bf k},\sigma }=E_{0}^{a}$ and introducing a local hybridization 
$V_{j,{\bf k},\sigma }=V_{j,\sigma }$, the eigenvalue problem of Eqs. (\ref
{E2.3}) or (\ref{E2.5}) has an exact solution,\cite{FoglioF} and the GFs can
be calculated analytically. As the problem is fully local, one can use the
Wannier representation for the creation and annihilation operators $C_{j{\bf %
,}\sigma }^{\dagger }$ and $C_{j,\sigma }$ of the c-electrons, and write $%
H_{r}=\sum_{j}H_{j}$, where $H_{j}$ is the local Hamiltonian 
\begin{equation}
H_{j}=\sum_{\sigma }\ \left\{ E_{0}^{a}\ C_{j{\bf ,}\sigma }^{\dagger
}C_{j,\sigma }+\ E_{j,\sigma }\ X_{j,\sigma \sigma }+\left( V_{j,\sigma }\
X_{j,0\sigma }^{\dagger }\ C_{j,\sigma }+V_{j,\sigma }^{\ast }\ C_{j{\bf ,}%
\sigma }^{\dagger }\ X_{j,0\sigma }\right) \right\} \qquad ,  \label{E2.9}
\end{equation}
and the subindex $j$ can be dropped because we assume a uniform system.

We shall denote with $\mid n,r\rangle $ the eigenstates of the Hamiltonian $%
H_{j}$ with eigenvalues $E_{n,r},$ where $n$ is the total number of
electrons in that state, and $r$ characterizes the different states. These
eigenstates satisfy 
\begin{equation}
{\cal H}\mid n,r\rangle =\varepsilon _{n,r}\mid n,r\rangle \qquad ,
\label{E2.10}
\end{equation}
\noindent where ${\cal H}$ is given in Eq. (\ref{E2.6}) and $\varepsilon
_{n,r}=E_{n,r}-n\mu $ (cf. Eq.~(\ref{E2.6a})). In Table \ref{T1} we give the
properties of the $\mid n,r\rangle $ states: number of electrons $n$, name
of the state $r$, $z$ component of spin $S_{z}$ and $\varepsilon
_{n,r}=E_{n,r}-n\mu $. The twelve eigenvalues $\varepsilon _{n,r}$ of ${\cal %
H}_{j}$ are represented in figure~\ref{F7_2}, and those corresponding to
different occupations $n=0,1,2,3$ are drawn in different columns. The states
are identified in the figure by the numbers $r$ above the levels, and the
lines joining different levels correspond to the possible transitions that
contribute to the GF. 

It is now straightforward to express the Fourier transform of the f-electron
GF in the form 
\begin{equation}
\left\langle \left( \widehat{X}_{j,0\sigma }(\omega _{s})\ \widehat{X}%
_{j,0\sigma }^{\dagger }(\omega _{s}^{\prime })\right) _{+}\right\rangle _{%
{\cal H}}=\Delta (\omega _{s}+\omega _{s}^{\prime })\ G_{ff,0\sigma
}^{at}(\omega _{s})\qquad ,  \label{E2.11}
\end{equation}
\noindent where 
\begin{equation}
G_{ff,0\sigma }^{at}(\omega _{s})=-e^{\beta \Omega }\sum_{n,r,r^{\prime }}%
\frac{\exp (-\beta \varepsilon _{n,r})+\exp (-\beta \varepsilon
_{n-1,r^{\prime }})}{i\omega _{s}+\varepsilon _{n-1,r^{\prime }}-\varepsilon
_{n,r}}\mid \langle n-1,r^{\prime }\mid X_{0\sigma }\mid n,r\rangle \mid ^{2}
\label{E2.12}
\end{equation}
and $\Omega =-kT\ln \sum \exp (-\beta \epsilon _{n,r})$ is the grand
canonical potential.\cite{Martinez} The equivalent equations for the
c-electrons are obtained by just replacing $\mid \langle n-1,r^{\prime }\mid
X_{0\sigma }\mid n,r\rangle \mid ^{2}$ in Eq.~(\ref{E2.12}) by $\mid \langle
n-1,r^{\prime }\mid C_{j,\sigma }\mid n,r\rangle \mid ^{2}$.

The f-electron GF can be written in the form

\begin{equation}
G_{ff,0\sigma }^{at}(\omega _{s})=-\exp (\beta \Omega )\ \sum_{j=1}^{8}\frac{%
m_{j}}{i\omega _{s}-u_{j}}\qquad ,  \label{E2.13}
\end{equation}
\noindent where $u_{j}$ are the poles and $m_{j}$ the residues of the GF.
There are only eight different $u_{j}$ for the f-electron GF, because
different transitions have the same energy and the residues of some
transitions are zero. Each $u_{j}=$ $-\left( \varepsilon _{n-1,r^{\prime
}}-\varepsilon _{n,r}\right) $ corresponds to the lines identified with $j$
that appear joining the levels in figure \ref{F7_2}, the two lines $u_{1}$
and the single lines $u_{3}$ and $u_{7}$ represent transitions that are
allowed in the absence of hybridization, while the remaining ones correspond
to transitions that are forbidden in that limit. It is important to notice
that for a system with given values of $E_{0}^{a}$, $E_{f}$ and $V$, the
position of the levels in figure \ref{F7_2} changes with the chemical
potential $\mu $. In that figure we have $E_{0}^{a}=\mu $ and $%
E_{f}<E_{0}^{a}=\mu $, and in that system the ground state is always the
singlet $\mid 2,9\rangle $, which has no magnetic moment in the absence of
field but can have a rather large induced moment because of the proximity of
the magnetic triplet.\cite{AlascioAA1}

\subsection{The atomic effective cumulant approximation}

\label{S02_5}

The atomic effective cumulant approximation (AECA) consists in substituting $%
M_{2,\sigma }^{eff}(z)$ in Eq.~(\ref{E2.8}) by an approximate $M_{2,\sigma
}^{at}(z)$ derived from the exact solution of the atomic limit, obtained by
solving for $M_{2,\sigma }^{at}(z)$ in the equation that is the atomic
equivalent of Eq.~(\ref{E2.8}). One then obtains 
\begin{equation}
M_{2,\sigma }^{at}(z)=\ \frac{\left( z-E_{0}^{a}+\mu \right) \ G_{ff,0\sigma
}^{at}(z)}{\left( z-E_{0}^{a}+\mu \right) -\mid V\mid ^{2}G_{ff,0\sigma
}^{at}(z)}\qquad ,  \label{E2.14}
\end{equation}
and from the point of view of the cumulant expansion, it contains all the
irreducible diagrams that contribute to the exact $M_{2,\sigma }^{eff}(z)$.
It should be emphasized that this diagrams contain loops of any size,
because there is no excluded site in this expansion, and all the filled
circles correspond to the same site, although they appear as different
vertices in the diagram. The difference between the exact and approximate
quantities is that different energies $E_{{\bf k},\sigma }$ appear in the
c-electron propagators of the effective cumulant $M_{2,\sigma }^{eff}(z)$,
while these energies are all equal to $E_{0}^{a}$ in $M_{2,\sigma }^{at}(z)$%
. Although $M_{2,\sigma }^{at}(z)$ is for that reason only an approximation,
it contains all the diagrams that should be present, and one would expect
that the corresponding GF would have fairly realistic features.

One still has to decide what value of $E_{0}^{a}$ should be taken. As the
most important region of the conduction electrons is the Fermi energy, we
shall use $E_{0}^{a}=\mu -\delta E_{0}$, leaving the freedom of small
changes $\delta E_{0}$ to adjust the results to particular situations, but
fixing its value for a given system when $\mu $ has to change to keep the
total number of electrons $N_{t}$ fixed, as for example when changing the
temperature $T$.

Another important point, is that concentrating all the conduction electrons
at $E_{0}^{a}$ would overestimate their contribution to the effective
cumulant, and we shall then reduce the hybridization by a coefficient that
gives the fraction of c-electrons which contribute most. We consider that
this is of the order of $V\varrho ^{0}$, where $\varrho ^{0}$ is the density
of states of the free c electrons per site and per spin, and to be more
definite we chose $\pi V\varrho ^{0}$, so the effective hybridization
constant $V_{a}$ coincides with the usual ``mixing strength'' $\Delta =\pi
V^{2}\varrho ^{0}$. This is essentially the same choice made by Alascio
et.al.\cite{AlascioAA} in their localized description of valence
fluctuations. Note that $V_{a}$ is only used in the calculation of $%
M_{2,\sigma }^{at}(z)$, and that the full value must be substituted in the $%
V $ that appears explicitly in Eq. (\ref{E2.8}), because the whole band of
conduction energies is used in $G_{c{,}\sigma }^{o}({\vec{k},}z)=-1/\left(
z-\varepsilon _{{\bf k\sigma }}\right) $.

The spectral density $\rho _{f}\equiv \rho _{f,0\sigma }(\omega )$ for
several values of $T$ is shown in figure \ref{F8_3} assuming the following
system's parameters: $E_{j,\sigma }=E_{f}=-0.5$, $\mu =0.$, a local
hybridization $V=0.3$ and a density of states of the unperturbed band
electrons given by a rectangular band of width $\pi $ centered at the
origin. As discussed before, we include adequate universal constants into
the different parameters, so that all of them are expressed in terms of a
single energy taken as unit. In the present case we can take that unit to be
equal to the total width of the band divided into $\pi $.

The localized energy $E_{f}$ of the local state is well below the Fermi
surface in this figure, corresponding to a typical Kondo region. The
spectral density $\rho _{f}$ does not change with $T$ below a certain value,
which is approximately $T=10^{-3}$ in figure \ref{F8_3}, because below this $%
T$ the state $\mid n=2,r=9\rangle $ is the only occupied in the atomic limit
and therefore $M_{2,\sigma }^{at}(z)$ does not change by further decreasing $%
T$. Other states are occupied at higher $T$, and the changes in $M_{2,\sigma
}^{at}(z)$ are reflected in $\rho _{f}$. The $\rho _{f}$ obtained by the
AECA at the lowest $T$ in the figure is basically the same $\rho _{f}$ of
the CHA in the region close to $\varepsilon _{f}=-0.5$, but it has also a
structure close to the Fermi surface (i.e. at $\omega =0$) that is absent in
the CHA. The spectral density in this region has the main characteristics of
the Kondo peak, namely its localization and the decrease of its intensity
when $T$ increases, as can be seen in figure \ref{F8_3}. This structure
shows a pseudo gap, as was obtained using other methods,\cite
{KagaKF,GreweSSC} but with a peak below the Fermi energy that is not as
sharp and that becomes more complex with increasing $T$, an effect that we
believe is due to the use of the atomic model to estimate $M_{2,\sigma
}^{eff}(z)$.

In figure \ref{F10_5} it is plotted the dependence of the spectral density
with $\mu $ showing a main structure that follows $\varepsilon _{f}$, and a
smaller one, corresponding to the Kondo peak, that remains fixed at the
Fermi surface when the system remains in the Kondo region, i.e. when $%
\varepsilon _{f}<0$. When $\varepsilon _{f}$ approaches the Fermi surface in
the intermediate valence region the two parts of the spectrum merge, and a
single structure then follows $\varepsilon _{f}$ for $\varepsilon _{f}>0$.

The AECA gives a $\rho _{f}$ that roughly agrees with the results obtained
by other methods,\cite{KagaKF} but the details of the spectral density near
the Kondo resonance depend in a very delicate way on the behavior of $%
M_{2,\sigma }^{eff}(z)$ near that region. In the calculation of $M_{2,\sigma
}^{at}(z)$ all the band structure is replaced by a single value $E_{0}^{a}$,
and it is therefore not surprising that the AECA results are less precise
than those of reference,\cite{KagaKF} which uses the decoupling of the
equations of motion followed by the self-consistent determination of the
averages resulting from that procedure. This last method, as well as the
dynamical mean field theory,\cite{Reviews} require some heavy computation
while the AECA is fairly simple from that point of view. As the spectral
density $\rho _{f}$ in the AECA has the same overall behavior shown by the
methods mentioned above, it can be used to obtain rather reasonable values
of many physical properties and of their dependence with $T$ and other
parameters. The main interest in this method, is that it is very natural to
make the extension to more complex systems with numerous local states, that
can be simplified by using the Hubbard operators to project the Hamiltonian
to the subspace of states of interest. The use of the AECA would then make
it possible to calculate properties of those systems without employing too
heavy computation. 


\section{Magnetic susceptibility and resistivity of FeSi}

\label{S03}

We discuss here the Kondo insulators within the context of the PAM,
employing the approximate GF introduced in Section \ref{S02_5}. We consider
the extreme case of $U\rightarrow \infty $, and apply it to {\rm FeSi},
which seems to behave like a typical Kondo insulator.\cite
{Schles,Aeppli,Schles2} Although this compound should be better described
with a finite $U$, we believe that this would not change too much the basic
properties from the ones presented here, but there would certainly be
changes in the parameters necessary to reproduce the measured properties of
FeSi when a finite $U$ is employed. Both the static conductivity $\sigma
(T)$ and magnetic susceptibility $\chi (T)$ vanish when $T\rightarrow 0$%
, and we can describe this dependence of the two properties, as well as
their initial rapid increase and posterior behavior for increasing $T$,\
with the same set of values of our model's parameters. The agreement with $%
\chi (T)$ is very good at not too high $T$, but to obtain a good agreement
at high $T$ one should consider that the parameters are affected by thermal
expansion in this region.

To select the total number $n_{tot}$ of electrons per site for our \ model
we first considered that in a band calculation of {\rm FeSi},\cite{Mattheis}
where all the {\rm Fe} $3d^{7}4s^{1}$ and {\rm Si} $3s^{2}3p^{2}$ states
were treated as valence electrons, the corresponding 48 valence electrons
per unit cell (4{\rm FeSi}) completely filled the lowest 24 valence bands.
To make a connection with the much simpler PAM, we first notice that a model
of two hybridized bands with two electrons per site would also fill the
lowest band, and that this model was considered to be a good starting point
to describe {\rm FeSi}.\cite{Aeppli} Therefore it seemed that $n_{tot}=2$
would be a good choice for our model. It is easier to keep a constant $\mu $
to calculate the $T$ dependence of properties, but the corresponding changes
in $n_{tot}$ where large enough to produce substantial changes in $\chi (T)$
but minor changes in $\rho (T)$ as the temperature increases. It was then
necessary to resort to the more laborious calculation at constant $n_{tot}$.

The parameters employed in the numerical calculations are numbers that have
to be scaled later to describe the different particular physical situations,
and for brevity we shall call them ``unscaled parameters''. This unscaled
parameters are usually chosen so that one physical magnitude, like the band
width, is equal to one. In our case we use a rectangular band, but fix the
unscaled total width $2W=5\ \pi $, this value (or any other of similar order
of magnitude) being convenient from the point of view of the numerical
calculation. The unscaled energy parameters can then be considered to be
given in units of $2W/5\pi $. In this paper we shall present a study
of a typical intermediate valence situation with $n_{tot}=2$, and these two
conditions are satisfied by employing values of $E_{f,\sigma }$\ and $\mu $
close to $4.$ when $2W=5\ \pi $. We shall then use as parameter the common
value $E_{f}$ of $E_{f,\sigma }$ in the absence of magnetic field, and fix
an unscaled value of $E_{f}=4.$ As a final step, the value of $\mu $ that
satisfies $n_{tot}=2$ has to be obtained before the $\chi (T)$ and $\rho (T)$
are calculated.

When $U\rightarrow \infty $, the average of the identity in the space of the
local states at site $j$ 
\begin{equation}
X_{j;\uparrow \uparrow }+X_{j;\downarrow \downarrow }+X_{j;00}=I_{j}\qquad ,
\label{E1.0}
\end{equation}
gives the ``completeness'' condition 
\begin{equation}
n_{f,\uparrow }+n_{f,\downarrow }+n_{f,0}=1\qquad ,  \label{E1.1}
\end{equation}
where $n_{f,\sigma }$ and $n_{f,0}$ are the average occupation number per
site of f-electrons states $\left| \ j,\sigma \right\rangle $ and $\left| \
j,0\right\rangle $ respectively. This relation gives the conservation of
probability of the local states at site $j$, and it is clear that it should
be satisfied before most properties of the system could be calculated. The
equivalent condition for the uncorrelated conduction electrons, obtained
from the identity $C_{j{\bf ,}\sigma }^{\dagger }C_{j,\sigma }+C_{j,\sigma
}C_{j{\bf ,}\sigma }^{\dagger }=1$, is satisfied for each $\sigma $ in all
the approximations that we have employed for the cumulant expansion, but the
Eq. (\ref{E1.1}) is not satisfied for the local electrons in the $%
U\rightarrow \infty $ limit.\cite{FFM} The origin of this problem has been
already discussed,\cite{FF3} and an exact solution for some families of
diagrams, together with a conjecture for the general case, has been
presented,\cite{FF2} but this treatment is not adequate for the approximate
GF we employ in this work. We have then applied an ``ad hoc''
renormalization, multiplying each unrenormalized $G_{ff,\sigma }^{u}({\bf k}%
,z)$ by a constant $x_{\sigma }$, so that the occupation numbers calculated
with the renormalized $G_{ff,\sigma }({\bf k},z)=x_{\sigma }G_{ff,\sigma
}^{u}({\bf k},z)$ satisfy Eq.(\ref{E1.1}). The two GF with $\sigma =\pm 1$
are equal in the absence of magnetic field, so that Eq.(\ref{E1.1}) is
sufficient to determine the two identical $x_{\sigma }$, and in the
following section we discuss how to obtain the two different $x_{\sigma }$
in the presence of a magnetic field.

\subsection{Magnetic susceptibility of FeSi}

The simplest way to calculate the static susceptibility is to consider the
one electron GFs in the presence of a small magnetic field in the $z$
direction, and obtain\ $n_{f,\sigma }$, $n_{c,\sigma }$ by employing Eqs.(%
\ref{E4.9}-\ref{E4.11}) and the corresponding relations for the conduction
electrons. The static magnetic susceptibility is then proportional to $%
g_{f}\ \left( n_{f,\downarrow }-n_{f,\uparrow }\right) +g_{c}\ \left(
n_{c,\downarrow }-n_{c,\uparrow }\right) $ when we neglect the Van Vleck
contributions, and the calculation of the corresponding approximate GFs in
the presence of a field is simpler when the gyromagnetic factors $g_{f}$ and 
$g_{c}$ of the two type of electrons are equal. The susceptibility $\chi (T)$
is then proportional to $n_{\downarrow }-n_{\uparrow }$ divided into the
magnetic field, where $n_{\sigma }=n_{f,\sigma }+n_{c,\sigma }$ is the total
number of electrons per site for each spin component $\sigma $. To calculate
the two factors $x_{\sigma }$ of the ``ad hoc'' renormalization we need of a
further relation besides Eq.(\ref{E1.1}). We notice that each $G_{ff,\sigma
}({\bf k},z)$ would give an independent $n_{f,0}$, and the two resulting
values are in general different in the presence of the field, so we ask of
them to be equal as the extra condition. The following symmetric equations
are then obtained 
\begin{eqnarray}
x_{\uparrow }\ D_{\uparrow }+x_{\downarrow }n_{\downarrow } &=&1\qquad , 
\nonumber \\
x_{\downarrow }\ D_{\downarrow }+x_{\uparrow }n_{\uparrow } &=&1\qquad ,
\label{E3.1}
\end{eqnarray}
where $n_{f,\sigma }$ and $D_{\sigma }$ are given in Eqs.(\ref{E4.9},\ref
{E4.14}).

To compare the ratio $\chi (T)/\chi (T_{m})$ with the corresponding
experimental value, it is necessary to chose the energy units to make the
fit, and it was found that a convenient method was to start the calculations
with the given unscaled $2W$ and $E_{f}$ and a trial set of the remaining
parameters, and then obtain the position $T_{m}^{calc}$ of the calculated
maximum of $\chi (T)$. The constant $s_{T}$ defined by $T_{m}=536\ K=s_{T}\
T_{m}^{calc}$ would then give the required scaling, so that multiplying all
the unscaled energy parameters times $s_{T}$, would give their absolute
values in units of degree $K$. As the maximum of $\chi (T)\ $is rather flat,
it was found that a better procedure was to find an $s_{T}$ that would give
the best fit of $\chi (T)/\chi (T_{m})$ at moderately low $T$. All the
adjustments were made by trial and error and direct comparison of the two
curves in the plot

A very good fit was obtained by keeping a constant $\mu =4.3$ with the $%
2W=5\pi $ and $E_{f}=4.$ discussed above, and \ the relation $n_{tot}=2.0$
was approximately satisfied at low $T$, but we could not find a similar set
of parameters that would give a good fit for all the values of $T$ when
requiring $\ $a constant $n_{tot}=2.0$. As the interval of temperatures is
rather large, we considered that because of the lattice thermal expansion,
some of the parameters could also change with $T$, and for simplicity we
employed a linear change in the hybridization constant: $V(T)=V_{0}\
(1+\alpha _{V}\ T)$, where it is sufficient to leave all the scaling in $%
V_{0}$ and use always the unscaled $\alpha _{V}$ and $T\ $in the product $%
\alpha _{V}\ T$. A very good fit shown in figure \ref{F7} was then obtained
with $V_{0}=1.8$ and $\alpha _{V}=-1.2$ and a position of the zero-width
conduction band, employed to calculate $M_{2,\sigma }^{at}(z)$, given by $%
E_{0}^{a}=\mu -\delta E_{0}$ with a $\delta E_{0}=0.3$ independent of $T$.
The scaling of the parameters was given by $s_{T}=1750\ K\simeq 0.1508\ $eV,
so that in absolute units $2W=2.379\ $eV, $E_{f}=603.2$\ meV, $V_{0}=271.4\ $%
meV and $\delta E_{0}=45.24\ $meV.

We shall employ the same values \ of these parameters when comparing the
theory with the measured resistivity in the following section, so that a
consistent description of the two properties be obtained.

It is possible from our calculations to analyze the changes with $T$ of the
total occupations $n_{f}$ and $n_{c}$ of the f and c electrons respectively,
and we have found that in the calculated curve of figure~\ref{F7} the $n_{f}$
changes from 0.47 at low $T$ to 0.53 at 800 K. These parameter values show
that the system remains in the intermediate valence region. It is also
interesting to compare the separate contributions $\chi _{f}(T)$ and $\chi
_{c}(T)$ of the f and c electrons to the total $\chi (T)=\chi _{f}(T)+\chi
_{c}(T)$, and these two quantities are plotted in figure \ref{F8}. Both
follow the same trend of the total $\chi (T)$, but $\chi _{c}(T)$ is much
smaller than $\chi _{f}(T)$, so that the choice of equal gyromagnetic
factors for the two type of electrons should not affect the final result in
a substantial way.

\subsection{Resistivity of FeSi}

\label{S03B}

The dynamic conductivity $\sigma \left( \omega ,T\right) $ is related to the
current current correlations by the well known Kubo formula.\cite{Kubo1,Mahan}
Two-particle GF are then necessary to calculate those correlations, and to
simplify the calculations, Schweitzer and Czycholl\cite{SchweitzerC}
employed the expression of the conductivity for dimension $d=\infty $ as an
approximation of the static conductivity for $d=3$. Only one-particle GFs
are then necessary to obtain $\sigma \left(  \omega ,T \right) $ in that limit,
because the vertex corrections cancel out,\cite{Khurana} and we shall follow
the same approach. As the hybridization is a hopping of electrons between
two different bands, it contributes to the current operator in the PAM,\cite
{CzychollL} but this contribution cancels out in our model because we employ
a local hybridization $V_{j,{\bf k},\sigma }=V_{j,\sigma }$. The expression
obtained contains explicit sums over ${\bf k}$, but it is possible to make a
further simplification by considering nearest-neighbor hopping in a simple
cubic lattice,\cite{MutouH,PruschkeCJ,ConsiglioG} and the sums over ${\bf k}$%
\ can be transformed\cite{MullerH} in integrals over the free conduction
electron energy $\varepsilon \left( {\bf k}\right) $. This transformation is
possible because in the AECA the $G_{cc,\sigma }({\bf k},\omega )$ is a
function of ${\bf k}$ only through the $\varepsilon ({\bf k})=\varepsilon $,
as both $M_{2,\sigma }^{at}(z)$ and $V_{j,{\bf k},\sigma }=V_{j,\sigma }$
are ${\bf k}$ independent. We can then write 
\begin{equation}
\rho _{c,\sigma }(\omega ;\varepsilon )=\frac{1}{\pi }\ \lim_{\eta
\rightarrow 0}\ {\it Im}\left\{ G_{cc,\sigma }({\bf k},\omega +i\left| \eta
\right| )\right\} \qquad ,  \label{E3.3}
\end{equation}
and the dynamical conductivity is given by 
\begin{equation}
\sigma \left(  \omega ,T\right) =C_{0}\frac{1}{\omega }\int_{-\infty }^{\infty
}d\omega \prime \left[ f_{T}(\omega \prime )-f_{T}(\omega \prime +\omega )%
\right] \int_{-\infty }^{\infty }d\varepsilon \ \rho _{c,\sigma }(\omega
\prime ;\varepsilon )\ \rho _{c,\sigma }(\omega \prime +\omega ;\varepsilon
)\ \varrho _{\sigma }^{0}(\varepsilon )\ \qquad ,  \label{E3.2}
\end{equation}
where $C_{0}$ is a constant discussed below and $\varrho _{\sigma
}^{0}(\varepsilon )$ is the density of states of the free conduction
electrons per site and per spin.

The static conductivity is then given by 
\begin{equation}
\sigma \left( T\right) =C_{0}\int_{-\infty }^{\infty }d\omega \ \left( -\ 
\frac{df_{T}(\omega )}{d\omega }\right) \ S(\omega )\ \qquad ,  \label{E3.4}
\end{equation}
where 
\begin{equation}
S(\omega )=\int_{-\infty }^{\infty }d\varepsilon \ \left( \rho _{c,\sigma
}(\omega ;\varepsilon )\right) ^{2}\ \varrho _{\sigma }^{0}(\varepsilon )\
\qquad .  \label{E3.4a}
\end{equation}

The constant 
\begin{equation}
C_{0}=\pi \ \frac{e^{2}}{\hbar }\ \frac{1}{a}\ \frac{2\ t^{2}}{d}\qquad ,
\label{E3.5}
\end{equation}
where $a$\ is the lattice parameter of FeSi and $t$ is the hopping parameter
of the hypercubic lattice that gives the c electron energy\cite{MullerH} 
\begin{equation}
\varepsilon ({\bf k})=-\left( \frac{2t}{\sqrt{2d}}\right) \sum\limits_{\nu
=1}^{d}\cos (a\ k_{\nu })\qquad .  \label{E3.6}
\end{equation}
We shall generally use a rectangular band with $-W\leq \varepsilon
\left( {\bf k}\right) $ $\leq W$, and to relate $t$ to $W$ we consider
that the band width \ $\sqrt{2\ d\ }2\ t$\ of Eq. (\ref{E3.6}) should be
equal to $2W$, so that $t=W/\sqrt{2\ d}$. Replacing $d=3$, $2W=5\pi $ and $%
a=4.489\AA $\cite{Mattheis} in Eq. (\ref{E3.5}) we find\cite{Note2} $%
C_{0}=6180/(\Omega \ cm)$.

If in the AECA of Eq. (\ref{E2.8b}) we abbreviate $G_{cc,\sigma }({\bf k}%
,z)=-\left[ a-\varepsilon ({\bf k})+i\ b\right] ^{-1}$, with real $a$ and $b$
(functions of $\omega $) defined by $a+i\ b=z+\mid V\mid ^{2}M_{2,\sigma
}^{at}(z)$, we find that when $b\rightarrow 0$ then $S(\omega )\sim {\rm %
O}\left| 1/b\right| $ if $\omega $ is outside the gap and $S(\omega
)\sim {\rm O}\left| b^{2}\right| $ if $\omega $ is inside. This
different behaviors can give rather different low $T$ limits of $\sigma
\left( T\right) $, because the integrand in Eq.(\ref{E3.4}) only contributes
in an interval of ${\rm O}\left| T\right| $ around the Fermi energy $\omega
=0$ (our frequency variables are given with respect to $\mu $). When $\omega
=0$ is inside the gap we then have a very small $\sigma \left( 0\right) $ if 
${\it Im\ }M_{2,\sigma }^{at}(0)\sim 0$, which corresponds to the case
of FeSi at low $T$, as is discussed below.

To analyze $M_{2,\sigma }^{at}(z)$ we notice from Eq. (\ref{E2.13}) that $%
G_{ff,0\sigma }^{at}(z)$ is real when $\eta \equiv {\it Im}\left[ z\right]
\rightarrow 0$, except at its only singularities on the real axis, that are
the poles at $z=u_{j}$. It is then clear from Eq. (\ref{E2.14}) that $%
M_{2,\sigma }^{at}(z)$ is real on the real axis of the $z$ complex plane,
\cite{Note3} except at the real solutions of $\left( \omega -E_{0}^{a}+\mu
\right) -\mid V\mid ^{2}G_{ff,0\sigma }^{at}(\omega )=0$, where it would
have poles with ${\it Im\ }M_{2,\sigma }^{at}(z)\neq 0$ in their
neighborhood. One should then make all the calculations at a finite $\eta $
and afterwards take $\eta \rightarrow 0$.

In figure \ref{F9} we plot the local spectral density of the conduction
electrons, namely 
\begin{equation}
\rho _{c,\sigma }(\omega )=\frac{1}{\pi }\ \lim_{\eta \rightarrow 0}\ {\it Im%
}\left\{ \frac{1}{N_{s}}\sum_{{\bf k}}G_{cc,\sigma }({\bf k},\omega +i\
\left| \eta \right| )\right\}
\end{equation}
as well as ${\it Im\ }M_{2,\sigma }^{at}(\omega +i\ \left| \eta \right| )$
as a function of $\omega $ for the same parameters employed in figure \ref
{F7}, but in this plot the variable $\omega $, as well as the fixed
parameters $T=0.001$ and $\eta =0.00001$, are given in the unscaled energy
units. It is clear from figure \ref{F9} that for these $\eta $ and $T$ the
peaks of ${\it Im\ }M_{2,\sigma }^{at}(\omega +i\ \left| \eta \right| )$ are
very sharp, and that the region where this quantity is appreciably large is
far away from $\omega =0$ in units of $T$. The value of the $b$ defined
above is therefore very small at that $\omega =0$, and the conductivity
would be extremely low because that value of $\omega $ is well inside the
gap so that $S(\omega )\sim {\rm O}\left| b^{2}\right| $. If the Fermi
surface ($\omega =0$ ) were inside the conduction band, the $S(\omega
)\sim {\rm O}\left| 1/b\right| $ and the conductivity would be extremely
large, and would tend to infinity when $\eta \rightarrow 0$. The physical
reason for this different behaviors at very low $T$ is the absence of
carriers when $\omega =0$ is inside the gap, and the absence of scattering
mechanisms for the c-electrons when $\omega =0$ is inside the band.

The extreme sharpness of the structure of ${\it Im\ }M_{2,\sigma }^{at}(z)$
is a consequence of the atomic approximation employed, and to alleviate this
character we have added an extra imaginary part $\eta _{a}=\left| \eta
_{a}\right| \ sgn({\it Im[z}])$ to its argument : $M_{2,\sigma
}^{at}(z)\Longrightarrow M_{2,\sigma }^{at}(\omega +i\eta _{a})$, so that
the poles of this quantity become Lorentzians that somehow mimic the effect
of the band width.To show the effect of this change, the plot of ${\it Im\ }%
M_{2,\sigma }^{at}(z)$ is also shown in figure \ref{F9} for the same
parameters used above but for an unscaled $\eta +\eta _{a}=0.001$. The ${\it %
Im\ }M_{2,\sigma }^{at}(z)$ has now an appreciable value at $\omega =0$ and
reaches inside the conduction band, and $\sigma (T)$ is much larger than
before, but it is still rather small because $T$ is small and only an
interval of order of $T$ contributes to $\sigma (T)$.

Addition of $\eta _{a}$ to the argument of $M_{2,\sigma }^{at}(z)$ leads to
similar effects as those already obtained by Mutou and Hirashima\cite{MutouH}
through ``introducing a small imaginary part $\Gamma $ to the conduction
electrons'', i.e. replacing $z=i\ \omega $ by $z+i\ \Gamma \ sgn(\omega )$
in the GFs $G_{ff,\sigma }({\bf k},z)$ and $G_{cc,\sigma }({\bf k},z)$.
Their justification is the existence in real systems of scattering processes
due to phonons and impurities, and we should also consider this mechanisms
as contributing to the $i\eta _{a}$. Within this interpretation one could
also consider a temperature dependence of $\eta _{a}$, but we have not
implemented this change in the present calculation.

It seems clear that the basic scattering mechanism in our calculation of the
PAM's $\sigma \left( T\right) $ is the hybridization, because the
otherwise free conduction electrons are scattered by the localized f
electrons through this interaction. This is apparent if we notice that the
relaxation effects are described by the imaginary part of the usual
self-energy $\Sigma _{cc,\sigma }({\bf k,}z)$, defined through 
\begin{equation}
G_{cc,\sigma }({\bf k},z)=-\left\{ z-\varepsilon ({\bf k})-\Sigma
_{cc,\sigma }({\bf k,}z)\right\} ^{-1}\qquad ,  \label{E4.1}
\end{equation}
and that the exact relation $\Sigma _{cc,\sigma }({\bf k,}z)=-\mid V({\bf k}%
)\mid ^{2}M_{2,\sigma }^{eff}({\bf k}{,}z)$ follows from Eq. (\ref{E2.8b}).
The relaxation mechanism of the c-electrons is then provided by the
hybridization, and the self energy is independent of ${\bf k}$ in the AECA: $%
\Sigma _{cc,\sigma }(z)=-\mid V\mid ^{2}M_{2,\sigma }^{at}(z)$.

As discussed above, the vanishing resistivity obtained in our approximation
when $\mu $ is inside the conduction band and $T\rightarrow 0$ is caused by
the atomic character of our effective cumulant $M_{2,\sigma }^{at}$, and the
introduction of a finite $\eta _{a}$ moderates this effect. In our
calculation of $\rho (T)=1/\sigma (T)$ we have chosen $\eta _{a}$ to
obtain a reasonable agreement with the measured values for $T$ above $50$ K,
because the conduction by ionized impurities would dominate $\sigma (T)$
at lower temperatures.\cite{MandrusSMTF} In figure \ref{F10} we plot $\rho
(T)$ for the same parameters employed to calculate $\chi (T)$ in figure 
\ref{F7} , but for an unscaled $\eta _{a}=0.0008$. The plot is compared to
the experimental values measured\cite{Schles} by Schlesinger et.al., which
we obtained by digitalization of their plot. We have preferred these values
to those of other authors\cite{Sales,Hunt} because the higher values of $%
\rho (T)$ seem to indicate a better sample with less impurities. The
agreement is fairly good above $50$ K, and the much lower experimental
resistivity below this $T$ is attributed to the ionic conduction, as
mentioned above. The impurity scattering is $T$ independent and could be
considered to be included in the value of $\eta _{a}$ we used, but the
phonon scattering would increase with $T$, and one should start with a
smaller $\eta _{a}$ and add a contribution that has the $T$ dependence
corresponding to this process. This effect could explain the difference
between the two curves above $T=200\,$\ K.

We point out that the typical values of $\eta _{a}$ we have used would not
affect the values of $\chi (T)$, although they were essential in fitting $%
\rho (T)$ to the experimental values.



\section{Conclusions}

\label{S04}

In this work we have used approximate\cite{Foglio} one-electron GFs of the
PAM in the limit $U\rightarrow \infty $ to study the static magnetic
susceptibility $\chi (T)$ and electrical resistivity $\rho (T)$ of FeSi,
which has the typical behavior of the Kondo insulators. The total number of
electrons per site $n_{tot}=2$ we have selected seems a good choice for
FeSi, and at each $T$ it is then necessary to find the value of the chemical
potential $\mu $ that gives this $n_{tot}$.

The exact GFs of the PAM are formally derived from their cumulant expansion, 
\cite{FFM} and are expressed in terms of an effective cumulant $M_{2,\sigma
}^{eff}({\bf k}{,}z)$, that is given by the contribution of all the proper
diagrams of the exact GF $G_{ff,\sigma }({\bf k},z)$. To obtain approximate
GFs, we substitute $M_{2,\sigma }^{eff}({\bf k}{,}$ $z)$, whose calculation
is as difficult as that of $G_{ff,\sigma }({\bf k},z)$, by the corresponding
quantity of an exactly soluble model, which is the same PAM but with a
conduction band of zero width. The use of this atomic limit makes the
corresponding effective cumulant $M_{2,\sigma }^{at}(z)$ independent of $%
{\bf k}$, and the approximate GFs have the same structure obtained by the
dynamical mean field theory.\cite{Kraut} While this last method uses
self-consistency conditions to obtain the final GFs, we use physical
considerations to chose the parameters that define $M_{2,\sigma }^{at}(z)$.
The approximation we employ does not give results as accurate as that
theory, but has the advantage that the numerical calculation is fairly
rapid, and that it can be extended to more complex systems with a large
number of local states, by projecting the large space of these states into a
manageable subspace of interest with the use of Hubbard operators.

We have shown that the spectral density $\rho _{f}$ of the approximate GF
employed in this work has, at low $T$ and in the Kondo region, a structure
on the Fermi surface that can be interpreted as the Kondo peak, a feature
that was missing in our cumulant expansions when only cumulants up to fourth
order were employed.\cite{FFM} This behavior is essential for the
calculation of the system's properties, and as the spectral densities
obtained by the AECA have the overall behavior that is expected from the
model,\cite{Reviews} the use of this approximation would then be expected to
give, without too heavy computation, rather reasonable values of many
physical properties and of their dependence with $T$ and other parameters.

These approximate one-electron GFs of the PAM in the limit $U\rightarrow
\infty $, are used here to describe the static magnetic susceptibility $\chi
(T)$ and electrical resistivity $\rho (T)$ of {\rm FeSi}, a compound which
has the typical behavior of the Kondo insulators. We have selected a total
number of electrons per site $n_{tot}=2$, because it seems a good choice for
FeSi, and at each $T$ it was then necessary to find the value of the
chemical potential $\mu $ that gives this $n_{tot}$. We tried to describe
the system employing a typical intermediate valence situation, and this was
achieved with a rectangular band of total width $2W=5\pi $ and an
unperturbed energy $E_{f}=4.$ of the f-electrons, requiring a chemical
potential $\mu \sim 4$ to satisfy $n_{tot}=2$. These unscaled values
correspond to measuring the energy parameters in units of $2W/5\pi $.

The susceptibility $\chi (T)$ is proportional to the difference $%
n_{\downarrow }-n_{\uparrow }$ between the occupation number of up and down
electrons, and we have fitted the calculated $\chi (T)/\chi (T_{m})$ to the
experimental results,\cite{Jacar} where $\chi (T_{m})$ is the maximum value
of the susceptibility. We have found that the contribution of the conduction
electrons to $\chi (T)$ is much smaller than that of the local electrons for
the employed parameters (cf. figure \ref{F8}), and the assumption made of
equal gyromagnetic factors for the two type of electrons has therefore no
major relevance.

To find the factor $s_{T\text{ }}$that transforms the unscaled parameters
into absolute units, we have scaled the calculated $\chi (T)/\chi (T_{m})$
curve so that it agrees with the experimental one at low $T$. It was not
possible to find a set of parameters that would give a good agreement at
both low and high $T$, and we considered that because of thermal expansion
some of the parameters could change with $T$. Assuming a linear dependence $%
V(T)=V_{0}\ (1+\alpha _{V}\ T)$ of the hybridization parameter $V$ it was
then possible to obtain a rather good agreement in the whole $T$ range (cf.
figure \ref{F7}). Although this is a very crude model of these effects and
one should also expect changes of $E_{f}$, it shows that the dependence of
the model's parameters with thermal expansion should not be neglected.

To calculate $\rho (T)$ we used an expression valid for infinite dimension, 
\cite{SchweitzerC} and assumed a nearest neighbor hopping in a simple cubic
lattice to transform the sums over ${\bf k}$ into integrals over the
unperturbed energies $\varepsilon ({\bf k})$ of the conduction electrons. As
a further non-essential simplification we used a rectangular band for the
corresponding spectral density $\varrho _{\sigma }^{0}$.

We have shown that the relaxation mechanism of the c-electrons is provided
by the hybridization, and the most likely situation in the AECA is that the
conductivity vanishes as $T\rightarrow 0$ when $\mu $ is inside the gap and
tends to infinity when $\mu $ is inside the band. Here we show that this
result is a consequence of the atomic character of the AECA, that makes $%
M_{2,\sigma }^{at}(z)$ imaginary only in the neighborhood of its finite
number of poles that are usually far apart from the Fermi surface ($\omega
=0 $). This behavior is moderated by introducing an extra imaginary part $%
\eta _{a}$ to its argument (cf. figure \ref{F9}), whose effect is similar to
that of scattering of the c-electrons by phonons and impurities.\cite{MutouH}%
\ The resistivity at low temperatures is probably determined by ionic
conduction, and if we choose $\eta _{a}$ so that the calculated resistivity
coincides with the experimental one close to $T\sim 50K$, a very good
agreement is obtained up to $T\sim 200K$.The departures at higher $T$
could be attributed to the phonon scattering.

In summary, we have shown that the spectral densities obtained by the AECA
have the overall behavior that is expected from the PAM, and we can
therefore expect that this approximation would give, without too heavy
computation, rather reasonable values of many physical properties and of
their dependence with $T$ and other parameters. We have employed the AECA to
study FeSi assuming a typical intermediate valence situation of the PAM in
the $U\rightarrow \infty $ limit, and we were able to give a good
description of both the static resistivity and magnetic susceptibility of
FeSi as a function of $T$ employing a single set of the model's parameters.

\section{Acknowledgements}

The authors are grateful to Profs. Roberto Luzzi, Marcelo J. Rozenberg,
Mucio A. Continentino and Sergio S. Makler for critical comments. They would
like to acknowledge financial support from the following agencies: CNPq
(MSF), FAPESP and CNPq (MEF). This work was done (in part) in the frame of
Associate Membership Programme of the International Centre for Theoretical
Physics, Trieste ITALY (MEF).







\begin{figure}[tbp]
\caption[Fig.1]{Typical cumulant diagrams for one-particle GF, where the
filled and empty circles represent the f-electron and c-electron free
propagators respectively, and the lines joining them represent the
hybridization interaction. (a) The diagrams of the chain approximation (CHA)
for the f-electrons, represented by the filled square to the right. (b) As
(a) but for the c-electrons, represented by an empty square. (c) A more
complicated diagram, with cumulants of fourth and sixth order. }
\label{F4_1}
\end{figure}

\begin{figure}[tbp]
\caption[Fig.2]{The energies $\protect\varepsilon _{n,r}=E_{n,r}-n\protect%
\mu $ of the twelve eigenstates $\mid n,r\rangle $ of the atomic limit are
represented in this figure, and those corresponding to different occupations 
$n=0,1,2,3$ are drawn in different columns. The index $r$ that characterize
the states is written above the corresponding levels, and the lines joining
different levels are identified by numbers $i$, showing the possible
transitions $u_i$ that contribute to the GF. As with Eqs.~(\ref{E2.6a},\ref%
{E2.6b}), the frequencies have the chemical potential $\protect\mu $
subtracted, so that the Fermi surface corresponds to $\protect\omega=0$. }
\label{F7_2}
\end{figure}

\begin{figure}[tbp]
\caption[Fig.3]{The spectral density $\protect\rho_{f}(\protect\omega )$ of
the f-electrons obtained with the AECA for several values of T, employing $z=%
\protect\omega +i\protect\eta $ with $\protect\eta=10^{-4}$. The system has
the following parameters: $E_{f}=-0.5$, $\protect\mu =0.$, $T=0.001$, a
local hybridization $V=0.3$ and a density of states of the unperturbed band
electrons given by a rectangular density of states of width $\protect\pi $
centered at the origin, all in the same energy units. The effective cumulant
was calculated employing a reduced hybridization $V_{a}=\Delta = V^2$ and a
frequency with a small imaginary part $\protect\eta _{a}=0.003$ (further
details of the introduction of $\protect\eta _{a}$ are given in
Subsection {\protect\ref{S03B}} ). }
\label{F8_3}
\end{figure}

\begin{figure}[tbp]
\caption[Fig.4]{The spectral density of the f-electrons for the same system
of figure~{\ref{F8_3}} but with fixed $T=0.001$ and for several values of
the chemical potential $\protect\mu $. It shows the crossover from the Kondo
region to the intermediate valence region. }
\label{F10_5}
\end{figure}

\begin{figure}[tbp]
\caption[Fig.7]{The reduced susceptibility $\protect\chi (T)/\protect\chi
(T_{m})$ as a function of $T$, where $\protect\chi (T_{m})$ is the maximum
value of $\protect\chi (T)$ . The open circles correspond to the
experimental values, taken from reference\onlinecite{Jacar}. The full line
corresponds to the model's calculation for a rectangular band with a total
width $2W=5\protect\pi $, an f electron energy of $E_{f}=4.$, a total number
of electrons $n_{tot}=2.0$, a $T$ dependent hybridization $V(T)=1.8(1.-1.2T)$
and the zero-width conduction band employed in the model located at $%
E_{0}^{a}=\protect\mu-\protect\delta E_{0}$ with a $T$-independent $\protect%
\delta E_{0}=0.3$. These unscaled values have to be multiplied by the scale
factor $s_T=1750\ K=150.8$ meV to express them in absolute units: $2W\simeq
2.379$ eV, $E_{f}=603.2$ meV, $V_{0}= 271.4$ meV and $\protect\delta E_{0}=
45.24$ meV. }
\label{F7}
\end{figure}

\begin{figure}[tbp]
\caption[Fig.8]{The separate susceptibilities $\protect\chi _f(T)$ (full
line) and $\protect\chi _c(T)$ (dotted line) of the f and c electrons
respectively, as a function of $T$ and for the same parameters of figure \ref
{F7}. }
\label{F8}
\end{figure}

\begin{figure}[tbp]
\caption[Fig.9]{The spectral density of the conduction electrons 
$\rho _{c,\sigma }(\omega )$
times $\pi $ (full line) and the imaginary part of the approximate effective
cumulant $M_{2,\sigma }^{at}(z)$ (dotted and dash-dotted lines)are plotted
in a logarithmic scale as a function of $\omega $ for $T=0.001$, $\eta
=0.00001$, $\eta _{a}=0.$ (full and dotted line), $\eta +\eta _{a}=0.001$
(dash-dotted line) and the same parameters of figure\ ~\ref{F7}. The values of 
$\omega $, $T$, $\eta $ and $\eta _{a}$ are given in unscaled energy units,
and have to be multiplied by the scale factor $s_{T}=1750$ K$=150.8$ meV to
express them in absolute units. }
\label{F9}
\end{figure}

\begin{figure}[tbp]
\caption[Fig.10]{The resistivity $\protect\varrho (T)$ in $\Omega $ cm,
measured by Schlesinger ed.al.\protect\cite{Schles} (triangles) and
calculated (full line) with Eq. (\ref{E3.4}) and the same parameters employed
in figure \ref{F7}, but with $\protect\eta _{a}=0.0008$. The values of $%
\protect\omega $ and $\protect\eta $ are given in unscaled units, and have
to be multiplied by the scale factor $s_T=1750$ K$=150.8$ meV to express
them in absolute units. }
\label{F10}
\end{figure}

\begin{table}[p]

\begin{center}

\begin{tabular}{|p{2em}|p{2em}|p{2em}|p{20em}|}
\hline
$n$ & $r$ & $S_{z}$ & $\varepsilon _{n,r}=E_{n,r}-n\mu $ \\ \hline
0 & 1 & $0$ & $E_{0}$ \\ \hline
1 & 2 & $+\frac{1}{2}$ & $\frac{1}{2}\left( E_{0}+E_{f}-\sqrt{\left(
E_{0}+E_{f}\right) ^{2}+4V^{2}}\right) -\mu $ \\ \hline
1 & 3 & $-\frac{1}{2}$ & $\frac{1}{2}\left( E_{0}+E_{f}-\sqrt{\left(
E_{0}+E_{f}\right) ^{2}+4V^{2}}\right) -\mu $ \\ \hline
1 & 4 & $+\frac{1}{2}$ & $\frac{1}{2}\left( E_{0}+E_{f}+\sqrt{\left(
E_{0}+E_{f}\right) ^{2}+4V^{2}}\right) -\mu $ \\ \hline
1 & 5 & $-\frac{1}{2}$ & $\frac{1}{2}\left( E_{0}+E_{f}+\sqrt{\left(
E_{0}+E_{f}\right) ^{2}+4V^{2}}\right) -\mu $ \\ \hline
2 & 6 & $+1$ & $E_{0}+E_{f}-2\mu $ \\ \hline
2 & 7 & $-1$ & $E_{0}+E_{f}-2\mu $ \\ \hline
2 & 8 & $0$ & $E_{0}+E_{f}-2\mu $ \\ \hline
2 & 9 & $0$ & $\frac{1}{2}\left( E_{0}+3E_{f}-\sqrt{\left(
E_{0}+E_{f}\right) ^{2}+8V^{2}}\right) -2\mu $ \\ \hline
2 & 10 & $0$ & $\frac{1}{2}\left( E_{0}+3E_{f}+\sqrt{\left(
E_{0}+E_{f}\right) ^{2}+8V^{2}}\right) -2\mu $ \\ \hline
3 & 11 & $+\frac{1}{2}$ & $E_{0}+2E_{f}-3\mu $ \\ \hline
3 & 12 & $-\frac{1}{2}$ & $E_{0}+2E_{f}-3\mu $ \\ \hline
\end{tabular}

\end{center}

\caption[TABLE I]{ The properties of the twelve eigenstates $\mid n,r\rangle
$ of ${\cal H}$ are given. The columns are labeled by the number of
electrons $n$, the name $r $ of the state, the $z$ spin component
$S_{z}$ of the state and the value of $\varepsilon
_{n,r}=E_{n,r}-n\mu $, where $ E_{n,r} $ is the energy of the
state $\mid n,r\rangle $. }
\label{T1}
\end{table}




\bigskip 

\begin{references}

\bibitem{Jacar}  V. Jacarino, G. K. Wertheim, J. H. Wernick, L. R. Walder
and S. Arajs, {\rm Phys. Rev.}{\it \ } {\bf 160}, 476 (1967)

\bibitem{Schles}  Z. Schlesinger, Z. Fisk, Hai-Tai Zhang, M. B. Maple, J. F.
DiTusa and G. Aeppli, {\rm Phys. Rev. Lett.} {\bf 71}, 1748 (1993)

\bibitem{Aeppli}  G. Aeppli and Z. Fisk, {\rm Comment. Cond. Mat. Phys.}{\it %
\ }{\bf 16}, 155 (1992)

\bibitem{Schles2}  Z. Schlesinger, Z. Fisk, Hai-Tai Zhang, M. B. Maple, {\rm %
Physica B}, {\bf 237-238}, 460 (1997)

\bibitem{Mattheis}  L. Mattheiss and D. Hamann, {\rm Phys. Rev.}{\it \ }{\rm %
B} {\bf 47}, 13114 (1993)

\bibitem{Fu}  C. Fu and S. Doniach, {\rm Phys. Rev. B} {\bf 51}, 17439 (1995)

\bibitem{Mucio}  M. A. Continentino, G. M. Japiassu and A. Troper,{\rm \
Phys. Rev. B} {\bf 49}, 4432 (1994)

\bibitem{TovarTOJC}  M. V. Tovar Costa, A. Troper, N. A. de Oliveira, G. M.
Japiassu and M. A. Continentino{\rm \ Phys. Rev. B} {\bf 57}, 6943 (1998)

\bibitem{Foglio}  M. E. Foglio, {\rm Brazilian Journal of Physics}{\it , }%
{\bf 27}, 644 (1997)

\bibitem{SanchezBC}  C. Sanchez-Castro, K. S. Bedell and B. R. Cooper,{\rm \
Phys. Rev. B} {\bf 47}, 6879 (1993)

\bibitem{Riseborough}  P. S. Riseborough,{\rm \ Phys. Rev. B} {\bf 45},
13984 (1992)

\bibitem{Varma}  C. M. Varma, {\rm Phys. Rev. B 50}, 9952 (1994)

\bibitem{Note1}  Varma\cite{Varma} argues that the absence of low energy
magnetic correlations, as observed in strongly correlated insulators, is
characteristic of the mixed valent character of these insulators. This could
be related to the lack of magnetic interactions between low energy
quasiparticles observed by neutron scattering experiments in 
FeSi.\cite{Aeppli,Tajima}

\bibitem{Hubbard4}  J. Hubbard, {\it \ {\rm Proc. R. Soc. London, Ser}. }%
{\rm A} {\bf 285}, 542 (1965)

\bibitem{HaleyErdos}  S. B. Haley and P. Erdos,{\it \ }{\rm Phys. Rev.} {\rm %
B }{\bf 5}, 1106 (1972)

\bibitem{Hubbard123}  J. Hubbard, {\rm Proc. R. Soc. London, Ser. A} {\bf 276%
}, 238 (1964)\newline
J. Hubbard, {\rm \ Proc. R. Soc. London, Ser. A} {\bf 277}, 237 (1964)%
\newline
J. Hubbard, {\it \ }{\rm Proc. R. Soc. London, Ser. A} {\bf 281}, 401 (1964)
(these are the first three papers of a series of six)

\bibitem{Hubbard5}  J. Hubbard, {\it \ {\rm Proc. R. Soc. London, Ser.} }%
{\rm A} {\bf 296}, 82 (1966)

\bibitem{FFM}  M. S. Figueira, M. E. Foglio and G. G. Martinez, {\rm Phys.
Rev. B} {\bf 50}, 17933 (1994)

\bibitem{Metzner}  W. Metzner, {\rm Phys. Rev. B} {\bf 43}, 8549 (1991)

\bibitem{Figueira}  M. S. Figueira, {\it Doctoral Thesis} (Campinas, SP,
Brasil: Universidade Estadual de Campinas, 1994)

\bibitem{Kraut}  A. Georges, G. Kotliar, W. Krauth and M. J. Rozenberg, {\rm %
Rev. Mod. Phys}{\it . }{\bf 68, }13 (1996)

\bibitem{MetznerV}  W. Metzner and D. Vollhardt,{\rm \ Phys. Rev. Lett.} 
{\bf 62}, 324 (1989)

\bibitem{MullerH}  E. Muller Hartmann, {\rm Z. Phys. B} {\bf 74}, 507 (1989)

\bibitem{Rozenberg}  M. J. Rozenberg, {\rm Phys. Rev. B} {\bf 52}, 7369
(1995)

\bibitem{RozenbergKK}  M. J. Rozenberg, G. Kotliar and H. Kajueter, {\rm %
Phys. Rev. B }{\bf 54}, 8452 (1996)

\bibitem{SchweitzerC}  H. Schweitzer and G. Czycholl, ${\rm Phys.\ Rev.\
Lett.}${\bf 67}, 3724 (1991)

\bibitem{Khurana}  A. Khurana, {\rm Phys. Rev. Lett.} {\bf 64}, 1990 (1990)

\bibitem{MutouH}  T, Mutou and D. S. Hirashima, {\rm J. Phys. Soc. Japan} 
{\bf 63}, 4475 (1994)

\bibitem{PruschkeCJ}  Th. Pruschke, D. L. Cox and M. Jarrell,{\rm \ Phys.
Rev.}{\it \ }{\rm B} {\bf 47}, 3553 (1993)

\bibitem{FetterW}  A. L. Fetter and J. D. Walecka, {\it Quantum Theory of
Many-Particle Systems} (McGraw-Hill, New York, 1971).

\bibitem{identity}  It is clear that the two sides of Eq. (\ref{E2.4}) give
the same result when applied to the basis \{$\left| \ j,0\right\rangle $,$%
\left| \ j,\downarrow \right\rangle $,$\left| \ j,\uparrow \right\rangle $,$%
\left| \ j,2\right\rangle $\} of the space of local states at site $j$.

\bibitem{Wortis}  M. Wortis, in {\it Phase Transitions and Critical
Phenomena }, edited by C. Domb and M. S. Green (Academic,London, 1974), Vol. 
{\bf 3}, pg.113.

\bibitem{Kubo}  R. Kubo,{\rm \ J. Phys. Soc. (Jpn.)}{\it \ }{\bf 17, }1100
(1962)

\bibitem{FFM2}  M. S. Figueira and M. E. Foglio,{\rm \ J.\ Phys.: Condens.
Matter} {\bf 8, }5017 (1996)

\bibitem{Hewson}  A. C. Hewson, {\rm J. Phys. C: Solid State Phys.} {\bf 10}%
, 4973 (1977)

\bibitem{YangW}  D. H. Y. Yang and Y. L. Wang, {\rm Phys. Rev. B} {\bf 10},
4714 (1975)

\bibitem{FF2}  M. S. Figueira and M. E. Foglio,{\it \ }{\rm J.\ Phys.:
Condens. Matter} {\bf 8, }5017 (1996)

\bibitem{Negele}  J. W. Negele and H. Orland {\it Quantum Many-Particle
Systems} (Addison-Wesley, New York, 1988), Chap. 2.

\bibitem{Zubarev}  D. N. Zubarev, {\rm Usp. Fiz. Nauk. }{\bf 71}, 71 (1960) [%
{\rm Sov. Phys.-Usp.} {\bf 3}, 320 (1960)]

\bibitem{Reviews}  A. C. Hewson, {\it The Kondo problem to Heavy Fermions (}%
Cambridge U.P. Cambridge, 1993)

P. Schlottmann, {\rm Phys.Rep.} {\bf 181}, 1 (1989)

P. Fulde, {\it \ }{\rm Solid State Physics }{\bf 41}, 1 (1988)

P. A. Lee, T. M. Rice, J. W. Serene, L. J. Sham and J. W. Wilkins,{\rm \
Comments Cond. Mat. Phys. }{\bf 12}, 99 (1986).

\bibitem{Kondo}  J. Kondo,{\rm \ Progr. Theor. Phys. }{\bf 32}, 37 (1964)

J. Kondo, {\rm Solid State Physics} {\bf 23}, 184 (1969)

\bibitem{Note01}  But note that the meaning of vertices and edges is
exchanged with that employed in the cumulant expansion.

\bibitem{LuttingerW}  J. M. Luttinger and J. C. Ward,{\rm \ Phys. Rev. }{\bf %
118}, 1417 (1960)

\bibitem{CracoG}  L. Craco and M. A. Gusm\~{a}o,{\rm \ Phys. Rev. B} {\bf 54}%
, 1629 (1996)

\bibitem{FFdi}  M. E. Foglio and M. S. Figueira, {\rm \ J.\ Phys. A
Mathematics and General }{\bf 30} 7879 (1997)

\bibitem{Atomic}  See Appendix A of Fulde's review and Appendix C of
Hewson's review, cited in reference \onlinecite{Reviews}.

\bibitem{FoglioF}  M. E. Foglio and L. M. Falicov, {\rm Phys. Rev. B} {\bf 20%
}, 4554 (1979)

\bibitem{FoglioBF}  M. E. Foglio, C. A. Balseiro and L. M. Falicov, {\rm %
Phys. Rev. B} {\bf 20}, 4560 (1979)

\bibitem{AlascioAA}  B. Alascio, R. Allub and A. A. Aligia, {\rm Z. Phys. B} 
{\bf 36, }37 (1979)

\bibitem{AlascioAA1}  A. Alascio, R. Allub and A. A. Aligia,{\rm \ J. Phys.
C: Solid State Phys. } {\bf 13}, 2869 (1980)

\bibitem{SimoesIA}  A. S. R. Sim\~{o}es, J. R. Iglesias and E. V. Anda,{\rm %
\ Phys. Rev. B} {\bf 29}, 3085 (1984)

\bibitem{Anda}  E. V. Anda,{\rm \ J. Phys. C: Solid State Phys.} {\bf 14},
L1037 (1981)

\bibitem{SimoesIRA}  A. S. R. Sim\~{o}es, J. R. Iglesias, A. Rojo and B.
Alascio, {\rm J. Phys. C: Solid State Phys.} {\bf 21}, 1941 (1988)

\bibitem{Simoes}  A. S. R. Sim\~{o}es, {\it Doctoral Thesis} (Porto Alegre,
RS, Brasil: Universidade Federal de Rio Grande do Sul, 1986 )

\bibitem{BrunetRI}  L. G. Brunet, R. Ribeiro-Teixeira and J. R. Iglesias, 
{\rm Sol. Stat. Comm.}{\it \ }{\bf 68}, 477 (1988)

\bibitem{BrunetGI}  L. G. Brunet, M. A. Gusm\~{a}o and J. R. Iglesias, {\rm %
Phys. Rev. B }{\bf 46}, 4520 (1992)

\bibitem{Consiglio}  R. Consiglio, {\it Doctoral Thesis} (Porto Alegre, RS,
Brasil: Universidade Federal do Rio Grande do Sul, 1997 )

\bibitem{Martinez}  G. G. Martinez Pino, {\it Doctoral Thesis} (Campinas,
SP, Brasil: Universidade Estadual de Campinas, 1989).

\bibitem{KagaKF}  H. Kaga, H. Kubo and T. Fujiwara,{\rm \ Phys. Rev. B }{\bf %
37}, 341 (1988)

\bibitem{GreweSSC}  N. Grewe, {\rm Sol. Stat. Comm. }{\bf 50}, 19 (1984)

\bibitem{FF3}  M. S. Figueira and M. E. Foglio,{\rm \ Int. J.\ Mod. Phys. B}%
{\bf 12, }837 (1998)

\bibitem{Kubo1}  R. Kubo,{\rm \ J. Phys. Soc. Japan}{\it \ }{\bf 12, }570
(1957)

\bibitem{Mahan}  G. D. Mahan {\it Many-Particle Physics} 
(Plenum Press, New York, 1990).

\bibitem{CzychollL}  G. Czycholl and H. J. Leder, {\rm Z. Phys. B} {\bf 44, }%
59 (1981)

\bibitem{ConsiglioG}  R. Consiglio and M. A. Gusm\~{a}o,{\rm \ Phys. Rev. B} 
{\bf 55}, 6825 (1997)


\bibitem{Note2}  The $G_{cc,\sigma }({\bf k},\omega )$ and $\rho _{c,\sigma
}(\omega ;\varepsilon )$ employed in Eq.~(\ref{E3.4}) are calculated with
the unscaled values of all the parameters, and in the calculation of $C_{0}$
one should then use the same type of units for $t^{2}$.

\bibitem{Note3}  The real value $M_{2,\sigma }^{at}(u_{j})=-\ ($ $%
u_{j}-E_{0}^{a}+\mu )/\mid V\mid ^{2}$ is taken at the poles $u_{j}$ of $%
G_{ff,0\sigma }^{at}(z)$.

\bibitem{MandrusSMTF}  D. Mandrus, J. L. Sarrao, A. Migliori, J. D. Thompson
and Z. Fisk,{\rm \ Phys. Rev. B} {\bf 51}, 4763 (1995)

\bibitem{Sales}  B. C. Sales, E. C. Jones, B. C. Chakoumakos, J. A
Fernandez-Baca, H. E. Harmon and J. W. Sharp,{\rm \ Phys. Rev. B} {\bf 50},
8207 (1994)

\bibitem{Hunt}  M. B. Hunt, M. A. Chernikov, E. Felder and H. R. Ott {\rm %
Phys. Rev. B} {\bf 50}, 14933 (1994)

\bibitem{Tajima}  K. Tajima, Y. Endoh, J. E. Fischer and G. Shirane,{\rm \
Phys. Rev. B} {\bf 44}, 6954 (1988)



\end{references}
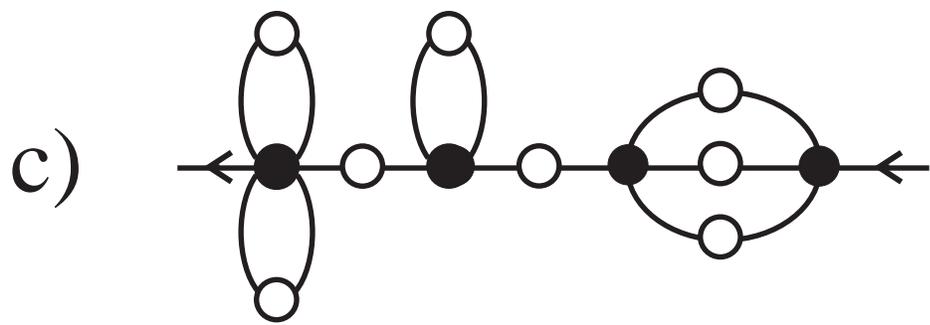
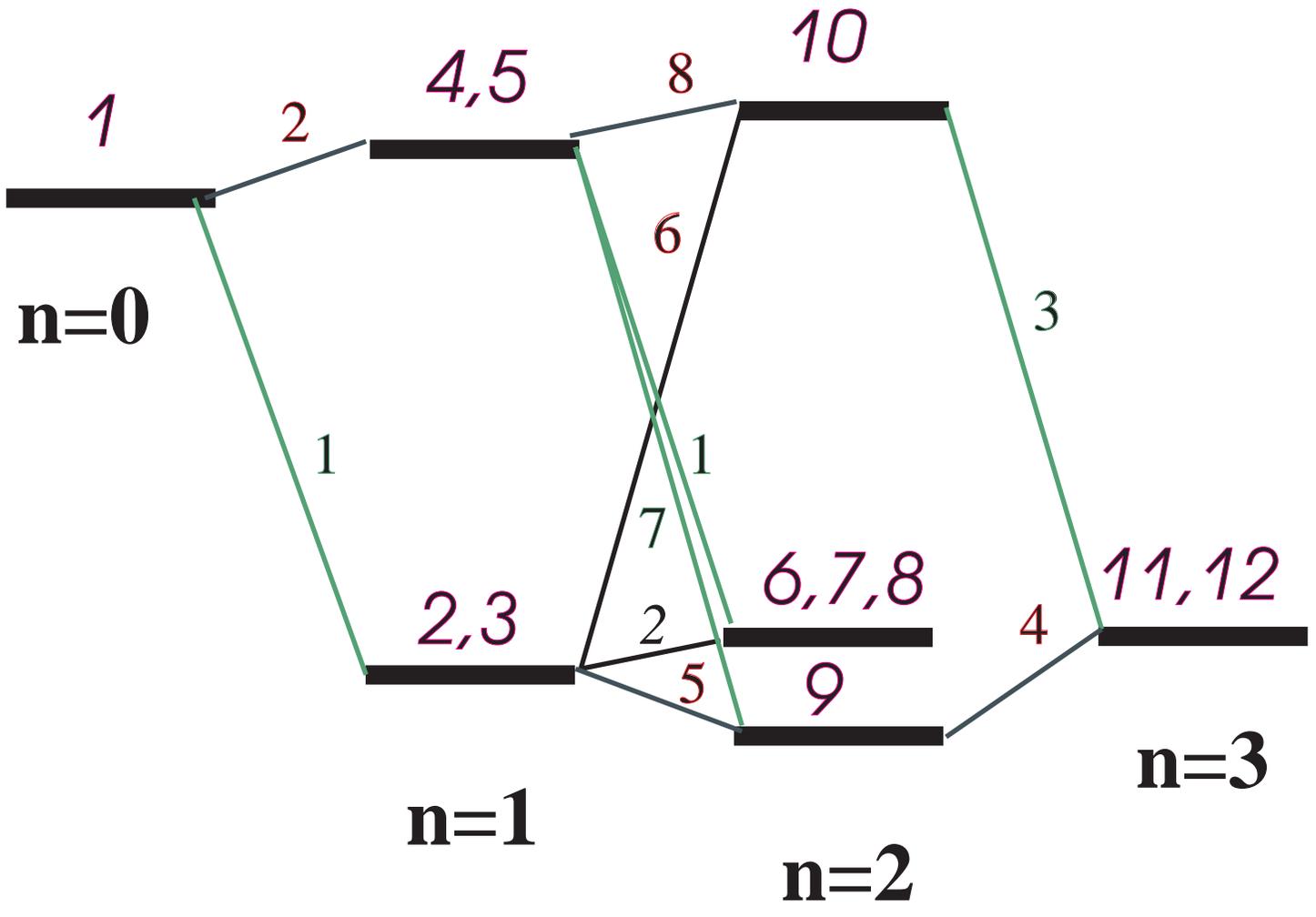
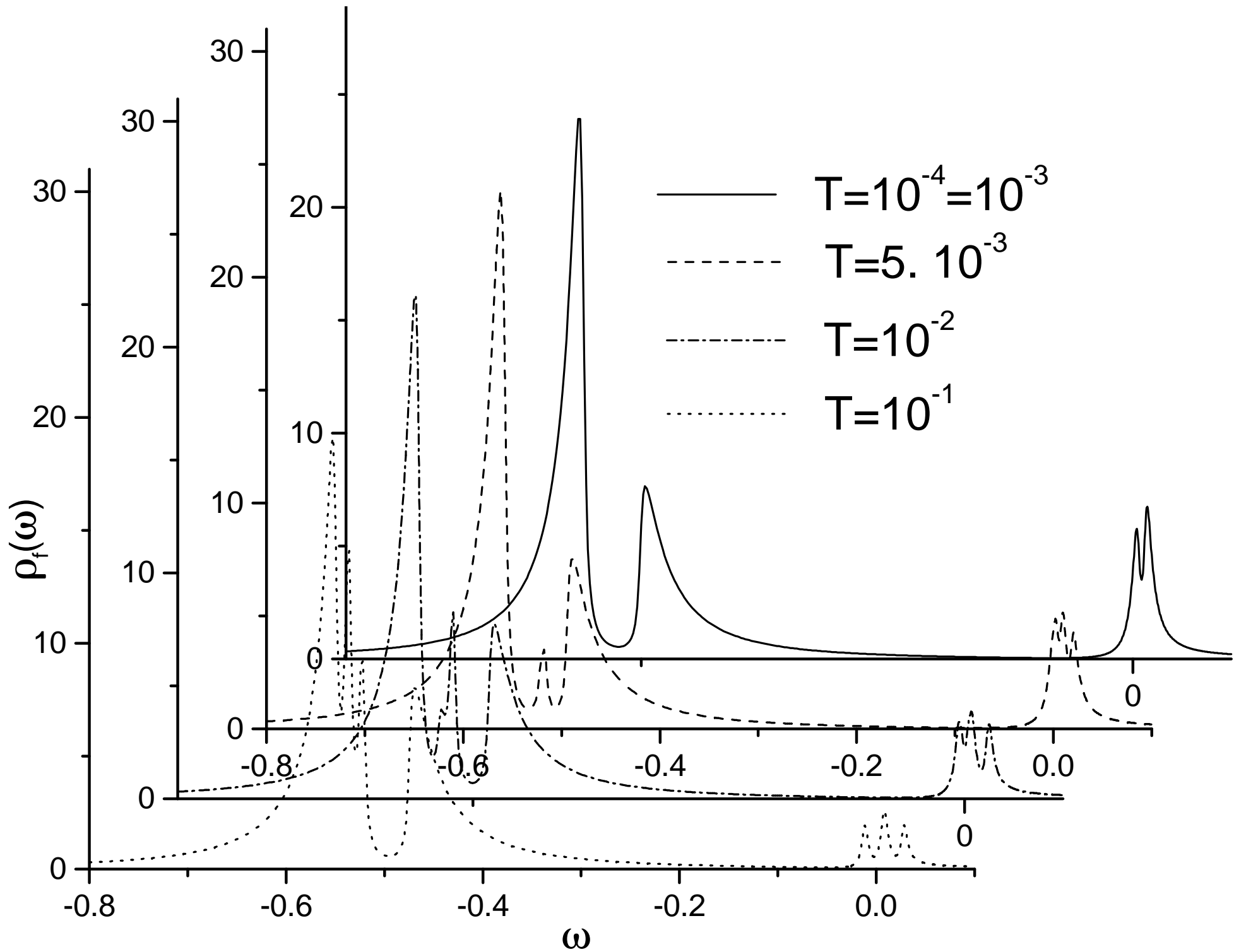
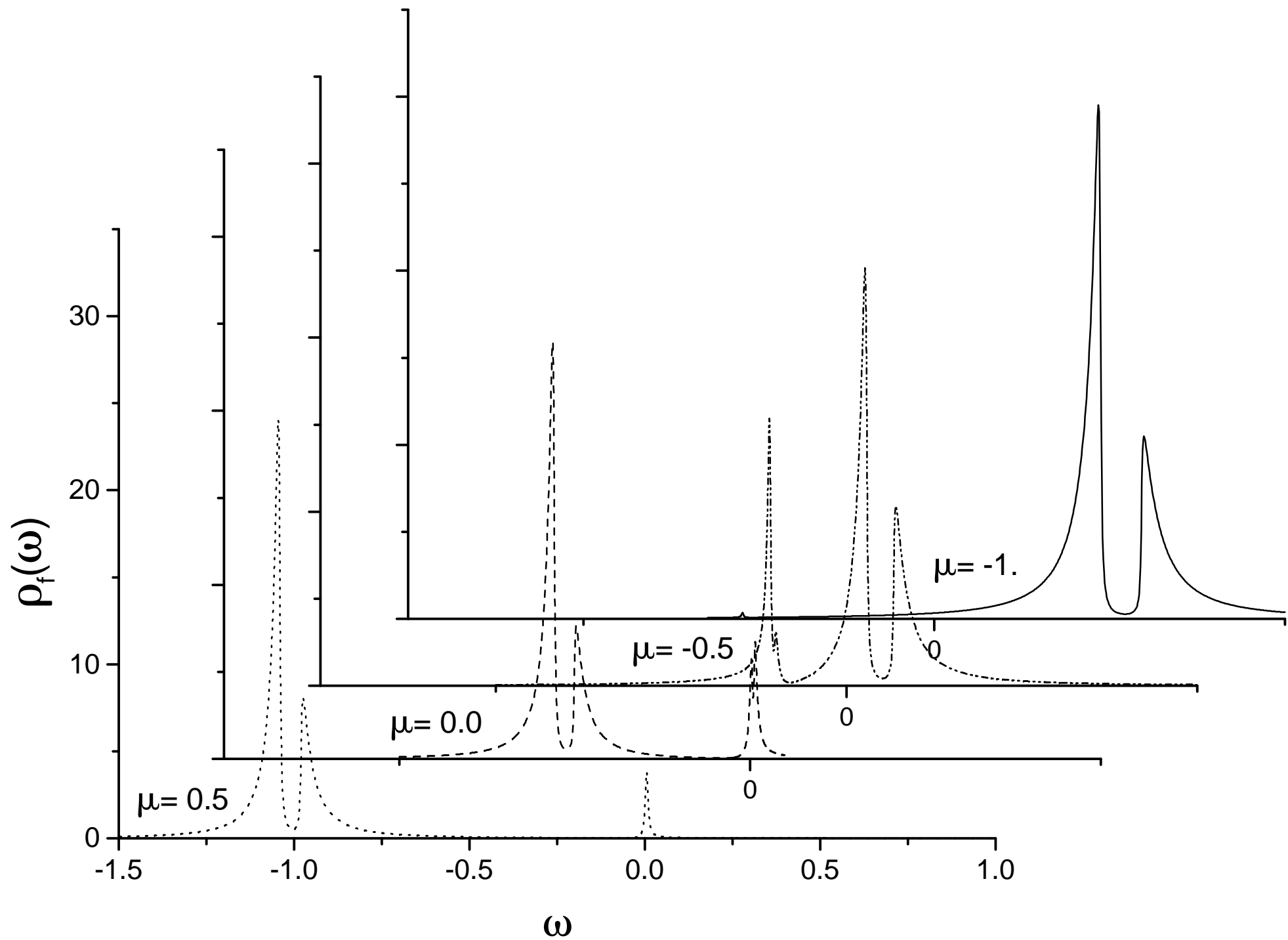
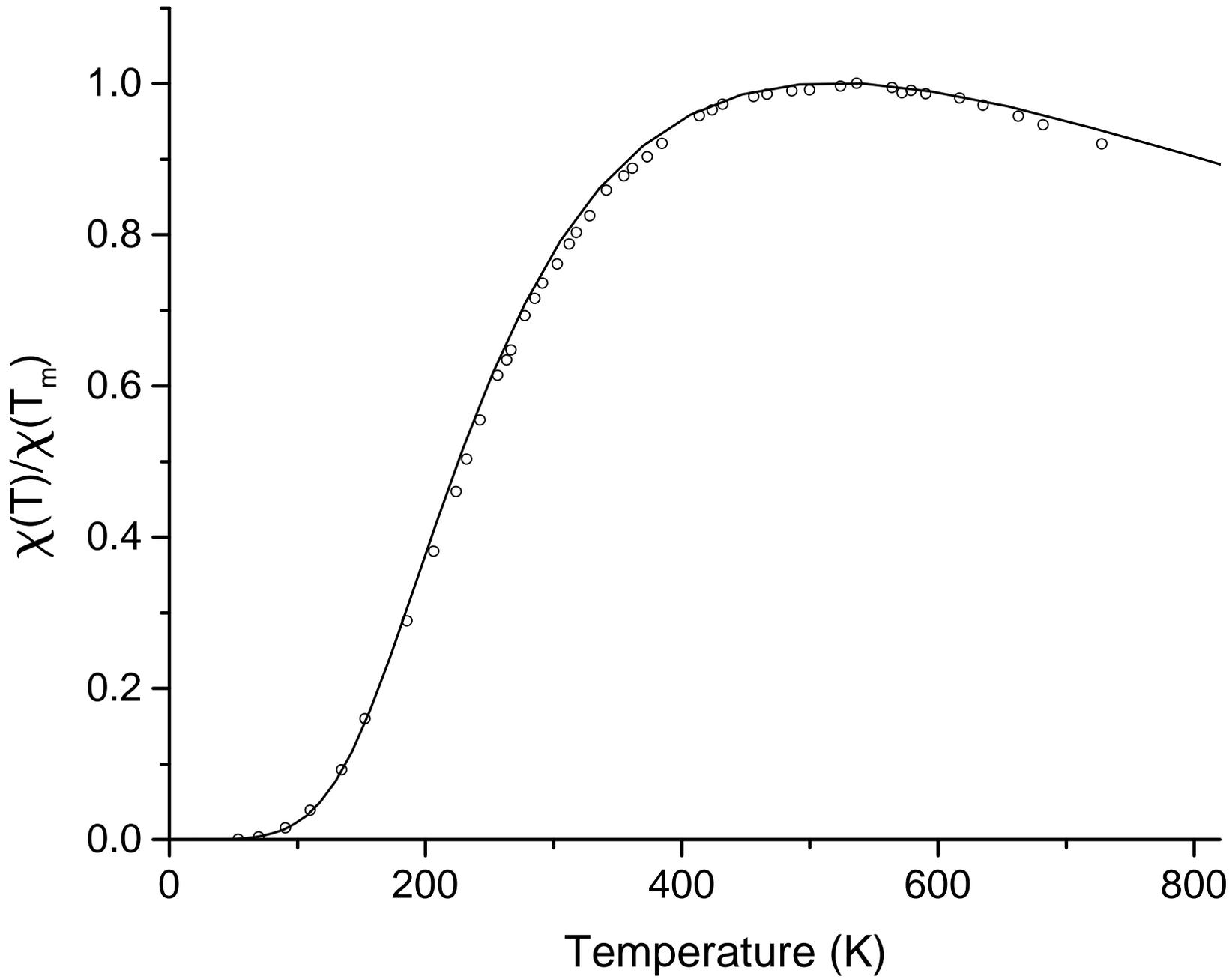
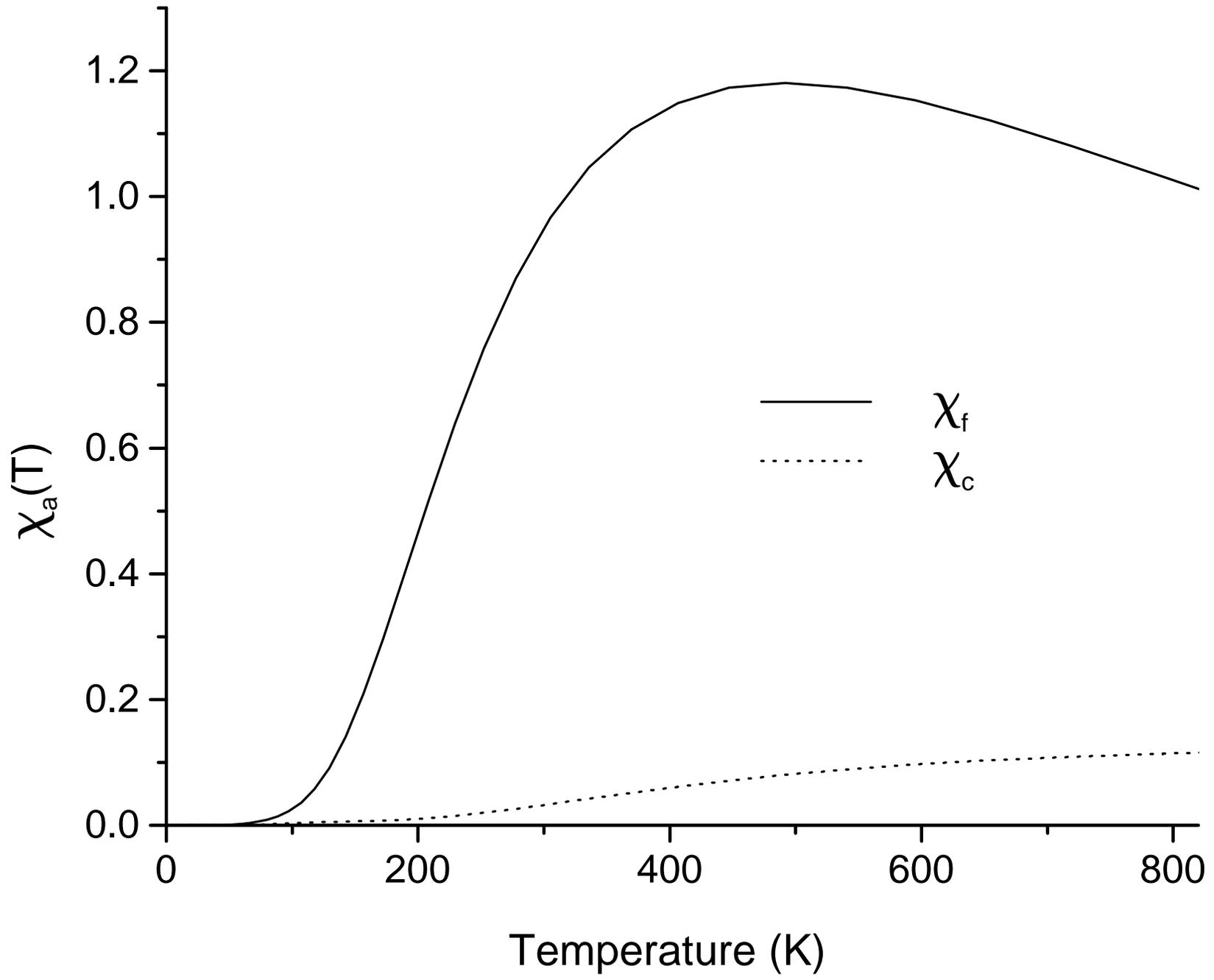
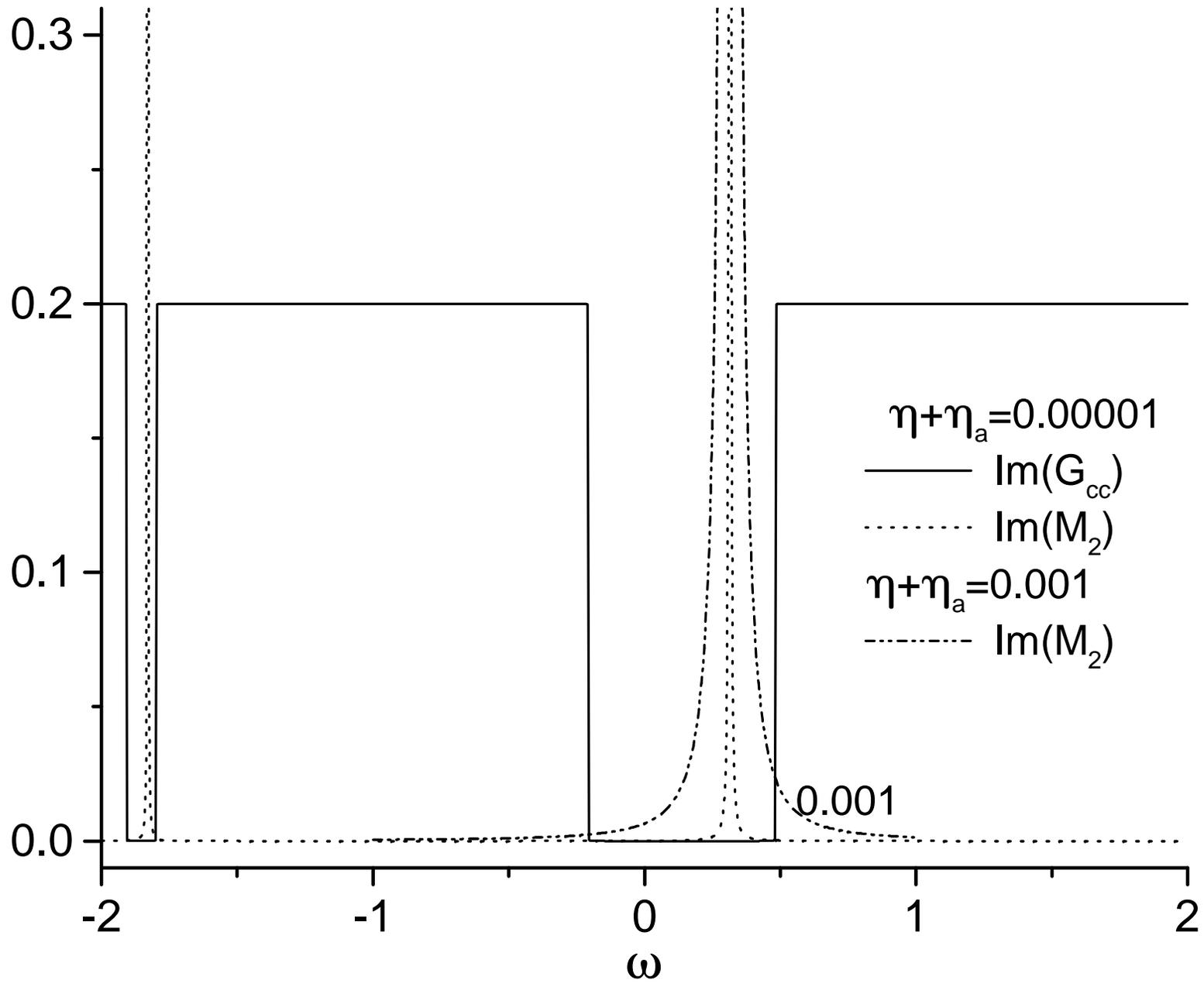
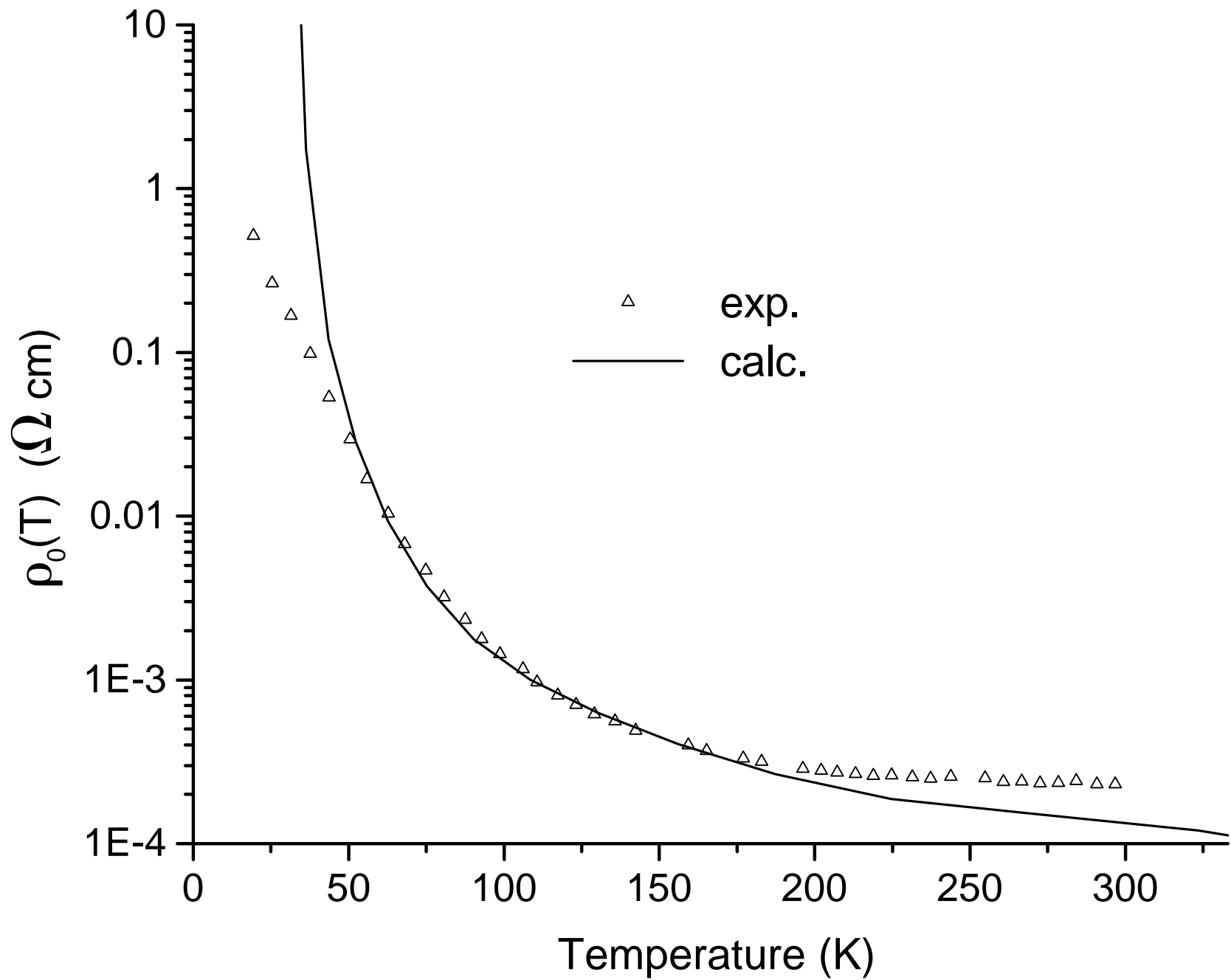
\end{document}